# KNIT user guide

I. Rychkova[2], V. Rychkov[1], K. Kazymyrenko[1,3], S. Borlenghi[1], X.Waintal[1,4]

1 Service de Physique de l'Etat condensé, DSM/DRECAM/SPEC, CEA Saclay 91191 Gif sur Yvette Cedex, France
2 Centre de Recherche en Informatique, Universite Paris1 – Pantheon – Sorbonne, 90 Rue de Tolbiac, 75013 Paris Cedex, France
3 Département Analyses Mécaniques et Acoustique, AMA/T64, EDF R&D, 1 avenue du Général de Gaulle 92141 Clamart Cedex, France
4 Service de Physique Statistique, Magnétisme et Supraconductivité, INAC/SPSMS, CEA 38054 Grenoble Cedex 9, France

***Abstract:*** *KNIT is a library that implements a fast and versatile algorithm to calculate local and global transport properties in mesoscopic quantum systems. Within the non equilibrium Green function formalism, KNIT applies a generalized recursive Green function technique to tackle multiterminal devices with arbitrary geometries. It is fully equivalent to the Landauer-Buttiker Scattering formalism. KNIT can (and has) been applied to many different tight-biding models for a large class of physical systems including topological insulators, graphene ribbons, hybrid superconducting systems and ferromagnetic multilayers.*

*KNIT main functionality is written in C++ and wrapped into Python, providing a simple and flexible interface for the user. Usual "input files" of numerical codes are replaced by small python scripts where the user builds the system and then compute whatever observable (conductance, noise, local density of states...) is needed. Python also provides simple graphical functions for pre and post treatments..*

*This paper is the KNIT documentation: it provides the information about KNIT installation and modeling principles that helps new users to start working with KNIT fast; it also provides a reference on KNIT functionality and describes the underlying algorithm, addressing more advanced users of KNIT. The KNIT algorithm is presented in details in the following paper:*

K. Kazymyrenko and X. Waintal "Knitting algorithm for calculating Green functions in quantum systems" Phys. Rev. B 77, 115119 (2008).

*The core developers of KNIT are Kyril Kazymyrenko and Xavier Waintal. This documentation has been written mainly by Irina Rychkova. Other developers include Valentin Rychkov, Simone Borlenghi and Havard Haugen.*

**KNIT is an open source project and is distributed freely. We only ask you to acknowledge the fact that you used KNIT and cite the main KNIT paper [ref [1], Phys. Rev. B 77, 115119 (2008)] in the publications where KNIT was used. To get the current version of KNIT, contact Xavier Waintal at: Xavier.Waintal@cea.fr**

## What you will find in this document:

1. Introduction: *Where does this KNIT code comes from. A few examples of calculations that have been performed with KNIT.*
2. KNIT quick tour: *a simple practical example and a brief theoretical introduction to the corresponding formalism.*
3. KNIT distribution: *what you will find in the various directories of the KNIT package.*
4. System requirements and installation: *how to make KNIT work on your machine? KNIT works very well on Linux. It has been seen to work well on Mac. If you want to compile it on Windows, you're on your own.*
5. Getting started with modeling: *A step by step explanation of how to build your own systems and calculate its transport properties. This section also explains the main functions needed to adapt the existing examples to your own systems.*
6. More advanced examples: *Advanced usage including modeling of graphene and how to optimize the calculation to save your computational resources.*
7. System functionality: *advanced usage: what other functions are available?*
8. System architecture: *advanced understanding: how the system is built?*

# 1. Introduction

Algorithms like the Recursive Green Function (RGF) algorithm have been around for several decades now. Those algorithms are based on a tight-biding description of a finite quantum system which is coupled to several semi-infinite electrodes. In its basic form, the RGF algorithm starts from a (left) electrode and recursively adds layers of the system until one reaches the second (right) electrode. Even this simplest form has proved to be very useful for simulating various mesoscopic systems (An early version written by Rodolfo Jalabert was referred in the community as "The Program"). As time went on and numerical simulations became more popular, more functionalities were needed. People started to develop new codes for other geometries, or to add more electrodes or internal degrees of freedom (to tackle say superconductivity or magnetism). In most cases each new problem was (is) associated with a new specific code. However, even though all these different problems can contain very different physics they share a common mathematical structure.

KNIT is an attempt to write a single program that allows addressing all these very different systems. In order to do that, we had to overcome two difficulties. First, we had to generalize the RGF algorithm in such a way that it did not rely on a specific geometry or topology of the model (This was done in Ref.[1]). Second, we needed a simple interface that allowed the user to easily build a new system without burying himself inside the code. Writing such an interface is very costly and they usually do not age well. Our technical choice here was to build KNIT as an extension of the Python language. Python is a programming language which is both extremely powerful and easy to learn and handle. An almost infinite number of modules have been written for Python so that only a few lines of codes are needed to send an email, perform a fast Fourier transform or produce a nice postscript file for a figure. KNIT is merely another module of Python. We believe that this programming paradigm (i.e. extending language capabilities as opposed to a standalone program) is very powerful in term of saving human time: codes are more versatile, age better and easily interact with other software.

Before going to the technical part of this documentation, we quickly show below a few examples of what has already been done with the code. Readers interested in the physics and the detailed of the modeling are referred to the corresponding scientific papers. Here we merely illustrate the sort of things that KNIT can do. Examples include:

- Calculations of the transport properties of an electronic Mach-Zehnder interferometer in a 2D gas in the Quantum Hall Regime (Fig.1, upper left)
- Anomalous Quantum Hall effect in a Hall bar made out of graphene nanoribbons (Fig.1, upper right)
- Calculation of the magneto-transport properties in a p-n junction made in a 2D topological insulator (CdTe/HgCdTe heterostructure). This work has been done in the group of Carlo Beenakker, see Ref.[2] (Fig.1, middle left)
- Spin accumulation in a magnetic nano-pillar (Fig.1 middle right)
- Cross Andreev Reflection in a graphene "Y" sample connected to 2 normal and 1 superconducting lead (group of Arne Brataas, see Ref.[6], Fig.1 lower left).
- Magnetic focussing and Cross Andreev Reflection in a 2DEG (group of Arne Brataas, see Ref.[7], Fig.1 lower right).

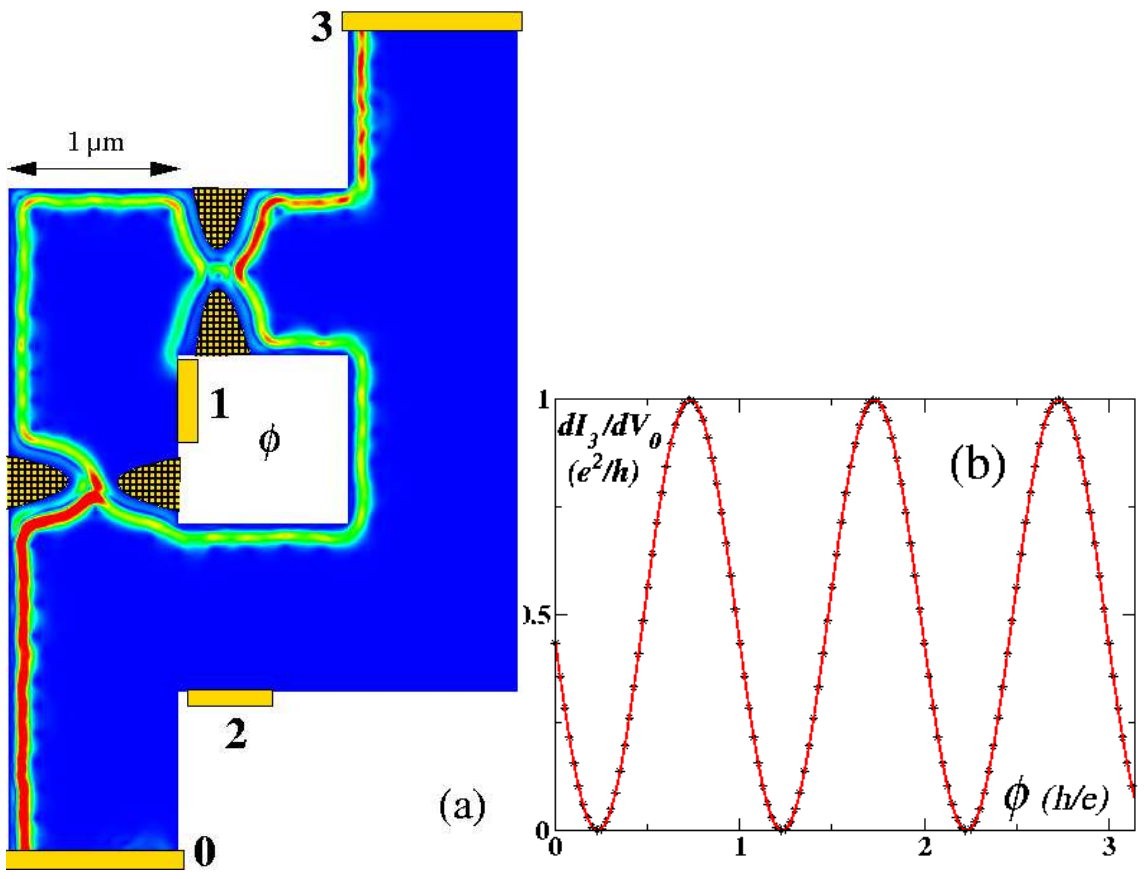

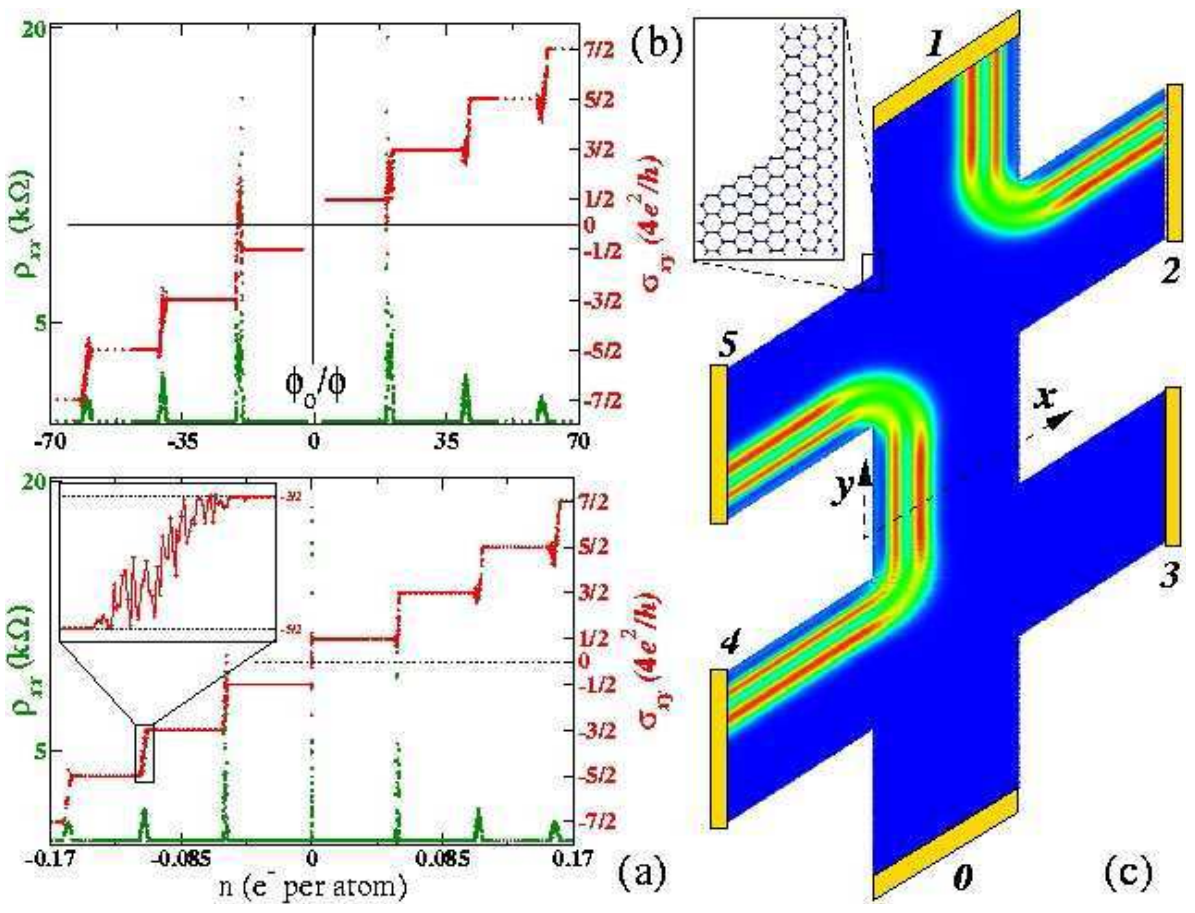

Here, we show a calculation of an electronic Mach-Zehnder interferometer in the quantum Hall regime where the 2D gas is modeled by a simple scalar tight-biding model on a square lattice. The first figure (a) illustrates the local current intensity when a bias voltage is applied to lead 0 and the other contacts are grounded (1.2 million sites, blue colors corresponds to no current and red to maximum current). One can observes the edge channel which is split by a first quantum point contact ("beam splitter") and then recombined by the second one. The second figure (b) shows the differential conductance between lead 0 and lead 3 as a function of the number of flux quanta through the hole. One finds the usual (cosine) interference pattern. See [Phys. Rev. B 77, 115119 (2008)] for more on this issue.

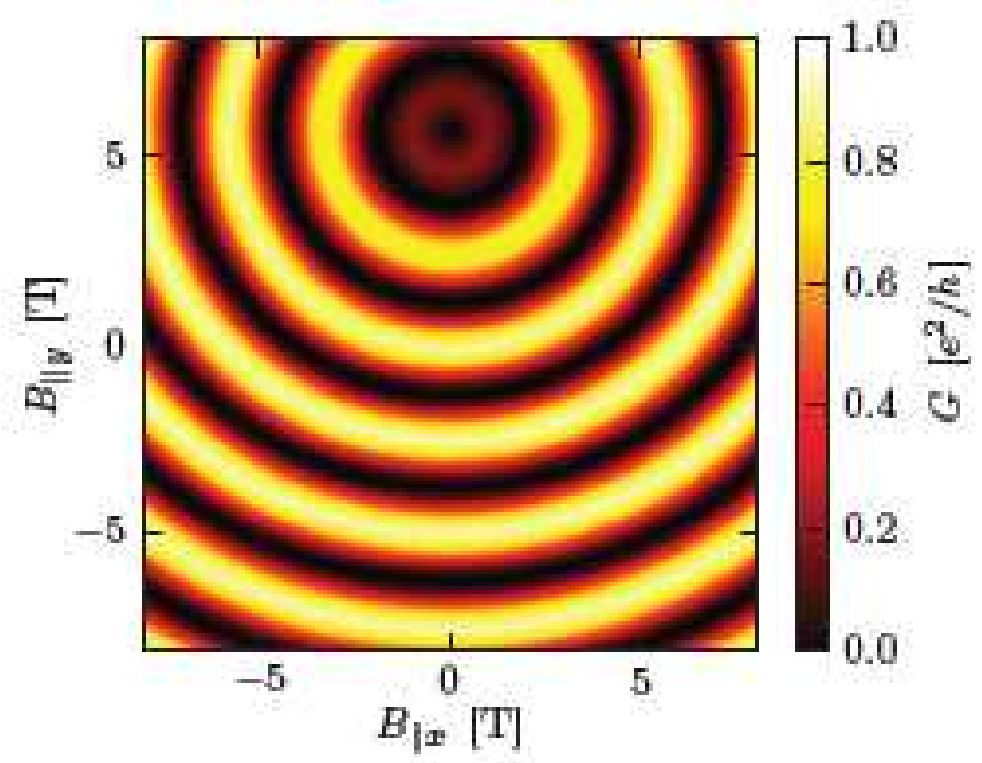

The following figure shows a calculation of a "H" shaped sample made out of a p-n junction of a topological insulator (HgTe/HgCdTe heterostrustures) in presence of a strong magnetic field. The plot shows a strong magneto-conductance as a function of a parallel magnetic field along the x and y directions ("Datta-Das" transistor). A tight-binding model with 4 orbitals per sites is used in this calculation. Calculation done in the group of Carlo Beenakker in Leiden. A. R. Akhmerov, C. W. Groth, J. Tworzydlo and C. W. J. Beenakker “Switching of electrical current by spin precession in the first Landau level of an inverted-gap semiconductor” Phys. Rev. B 80, 195320 (2009).

This is a calculation of the Quantum Hall effect in a graphene Hall bar (with zigzag nanoribbons). The Hall conductance $\sigma_{xy}$ and longitudinal resistance $\rho_{xx}$ are plotted as a function of inverse magnetic field (a) and carrier density (b) in presence of a small disordered potential (10% of the hopping matrix elements). The anomalous quantum Hall effect specific of graphene is nicely recovered. The inset of (b) shows a zoom of the transition between plateaus. Figure (c) shows the local current intensity when current is injected from both contact 1 and 4 which allows to study the edge channels. See [Phys. Rev. B 77, 115119 (2008)] for more on this issue.

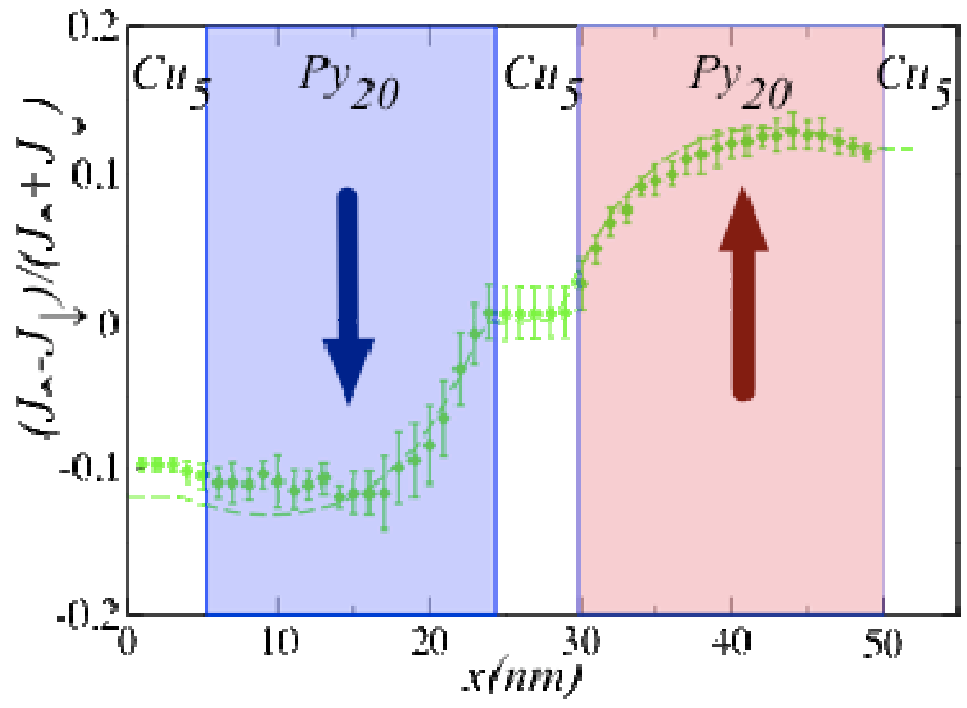

This figure shows the spin accumulation profile of a magnetic nanopillar (spin valve) made of the following stack: Copper (5 nm), Permalloy (20 nm), Copper (5 nm), Permalloy (20 nm) and Copper (5nm). The symbols correspond to the calculations done using KNIT with a spin dependent tight-binding model (the error bar corresponds to the average over different symbols) while the dashed line corresponds to a Random Matrix Theory aproach. (results *not yet published).*

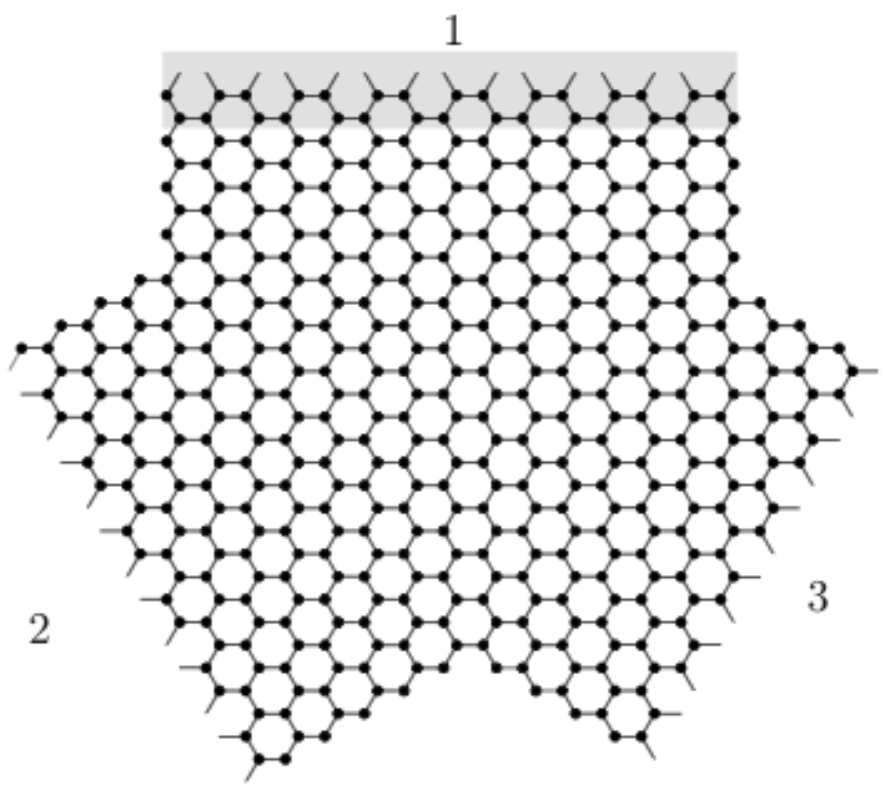

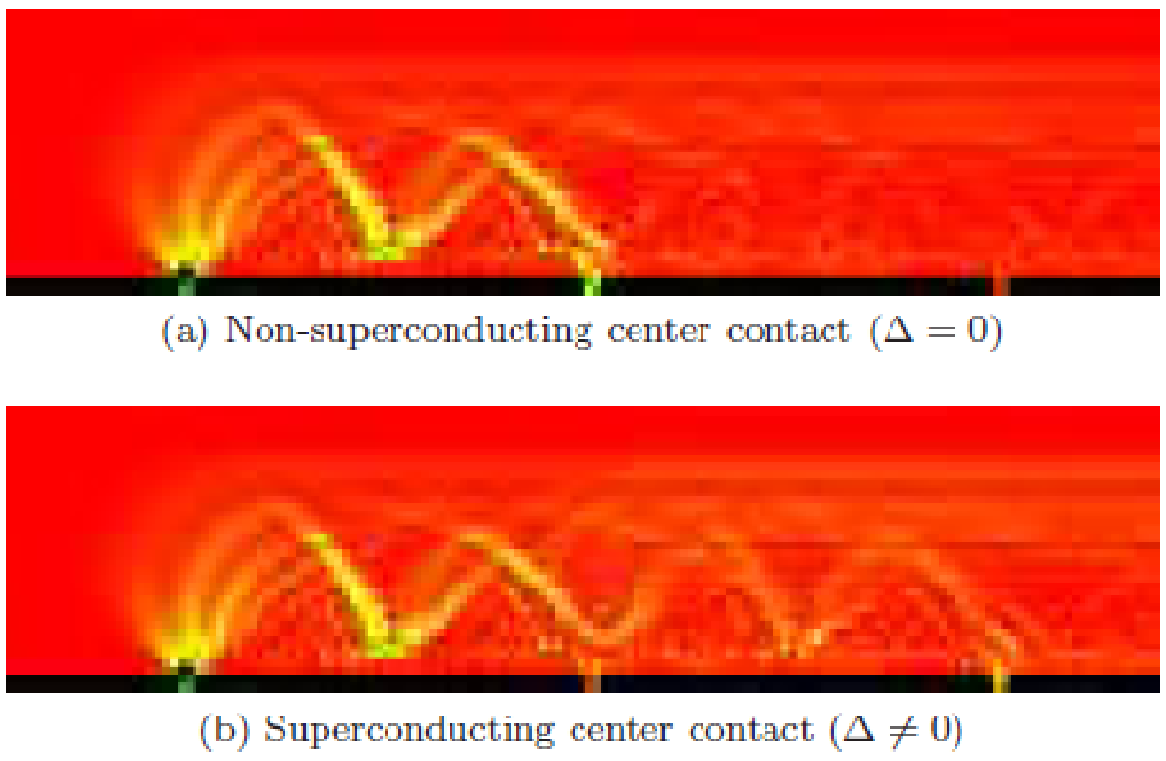

Calculations of Cross Andreev Reflection in a graphene "Y" shaped sample connected to 2 normal and 1 superconducting electrode. Group of Arne Brataas, Trondheim. "Crossed Andreev reflection versus electron transfer in three-terminal graphene devices", Havard Haugen, Daniel Huertas-Hernando, Arne Brataas, and Xavier Waintal Phys. Rev. B **81**, 174523 (2010).

This calculation shows Cross Andreev Reflection magnified by magnetic focussing. The tight-biding model includes an electron/hole grading to account for superconductivity. Calculation done in the group of Arne Brataas, Trondheim. "Focused Crossed Andreev Reflection", Havard Haugen, Arne Brataas, Xavier Waintal and Gerrit E. W. Bauer, arXiv:1007.4653.

FIG. 1 Examples of using KNIT for modelling mesoscopic systems.

KNIT computing kernel is written entirely in C++ except for the (external) fortran BLAS library that takes care of the linear algebra. An automatic wrapping technique is used to produce the interface with the PYTHON programming language providing fast development and many post and pre-processing tools. No C/C++ knowledge is needed to use KNIT.

This document contains some material to get started with KNIT. We hope that this tool will be useful to other researchers working in mesoscopic theory or experiment or even for educational purposes.

# 2 KNIT quick tour

## *2.1 What exactly KNIT can do for you?*

The input of a KNIT calculation is a quantum system connected to electrodes. In practice, one has to provide a tight-biding model that describes the finite quantum region of interest as well as a description of the electrodes. The latter are semi-infinite periodic systems. The bare output of KNIT are various elements of the retarded Green function of the system. Although several observables (like the conductance) can be obtained directly, one has to fall back to the retarded Green function anytime a new observable is needed. KNIT calculates only a subpart of all Green function elements:

a) between the leads (to study global transport properties of the system, such as conductance, Shot noise etc);

b) at a given site inside the system (to study single particle density)

c) between neighbouring sites inside the system (to study local current density)

As for b) and c), KNIT calculates the Retarded and the lesser (i.e. non equilibrium) Green function of the system. If you don't know much about Green functions, you can still use KNIT to calculate a few observables (conductance mainly) but will need to get acquainted with them to proceed further.

## *2.2 A first practical Example:*

The code listing below is a simple script written in Python programming language. This is an example of input instructions for KNIT.

```
1.  t=knit.scalarM(1.)
2.  V=knit.scalarM(0.)
3.  E=-0.5
4.  MyA=knit.rectangle([8,4],[0,3],t,v)
5.  MyB=knit.rectangle([3,7],[5,0],t,v)
6.  MyA.coller(MyB)
7.
8.  interface_contact1=knit.rectangle([1,4],[0,3],t,V)
9.  interface_contact2=knit.rectangle([3,1],[5,0],t,V)
10. lead1=knit.unreservoirN(interface_contact1,t)
11. lead2=knit.unreservoirN(interface_contact2,t)
12.
13. MySystemWithLeads=knit.systemtotal(MyA,lead1)
14. MySystemWithLeads.addlead(lead2)
15. system.visu2D(MySystemWithLeads, s+".MySystem")
16.
17. for i in range(10):
18.     E=(float(i)/10)*2.-1.99
19.     G_MATRIX=obs.conductance_matrix(MySystemWithLeads,E)
20.     print E,G_MATRIX[1,0]
```

Hopping matrix (line 1); Onsite energy matrix (line 2); Size of the block: 8x4; Position of the block: x=0; y=3 (line 4); Gluing B on top of A (line 6); Interface parts as rectangular blocks.. (lines 8-9); System with 1 lead (line 13); Adding the second lead (line 14); Visualization (line 15); Calculation.. (line 19); Printing results on the screen (line 20)

This 20 lines script does the following:

1. Defines the hopping matrix **t**, the on-site energy matrix **V**, and the energy **E** at which transport through our mesoscopic system is studied (lines 1-3);
2. Defines the building blocks A and B for a system and constructs the system of a required geometry as illustrated in Fig. 2-a (lines 4-6);

3. Creates interface contacts between 2 leads and the systen then leads themselves (lines 8-11)
4. Creates a total system (Fig.2-b) by attaching the leads to a system defined in the step 2 (lines 13-14), plots the resulting system, saving the image in *.eps, and *.pdf files (line 15)
5. Calculates the conductance matrix of the constructed system depending on the energy and prints the results for 10 different values of energy E on the screen (Fig.3).

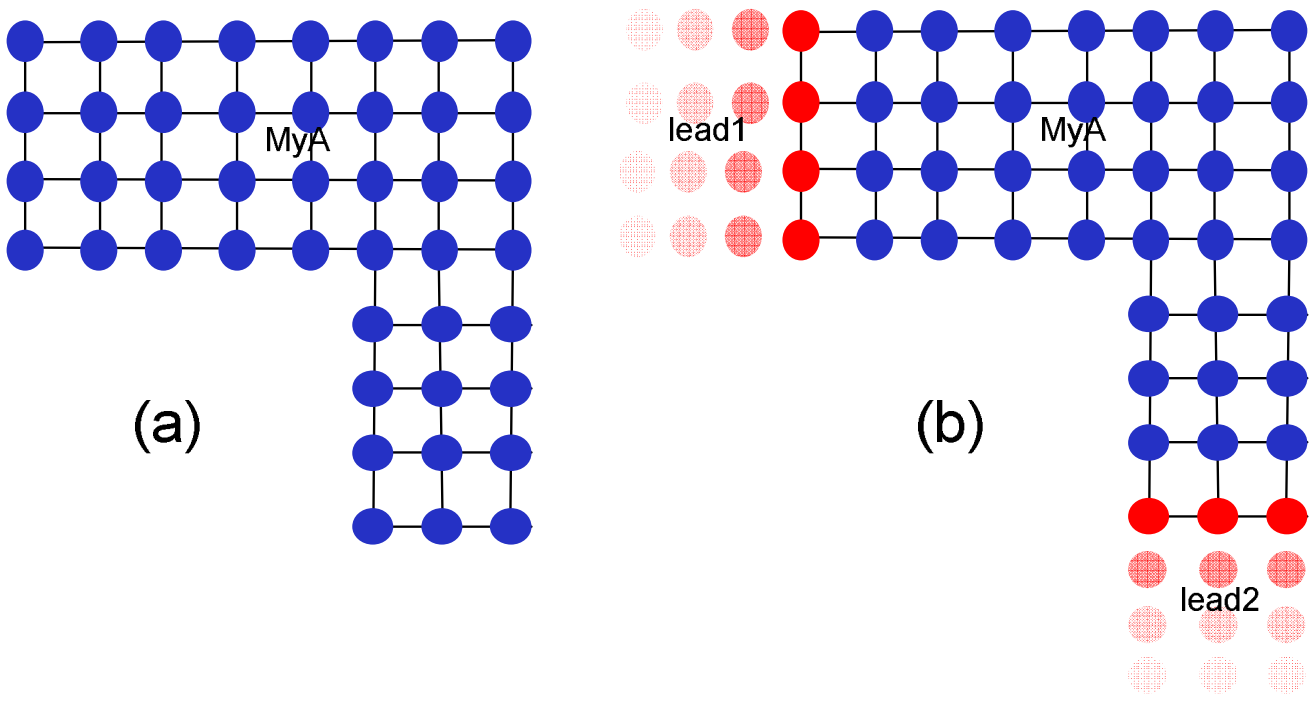

**FIG.2:** a) System MyA constructed by gluing 2 blocks; b) Total system with leads

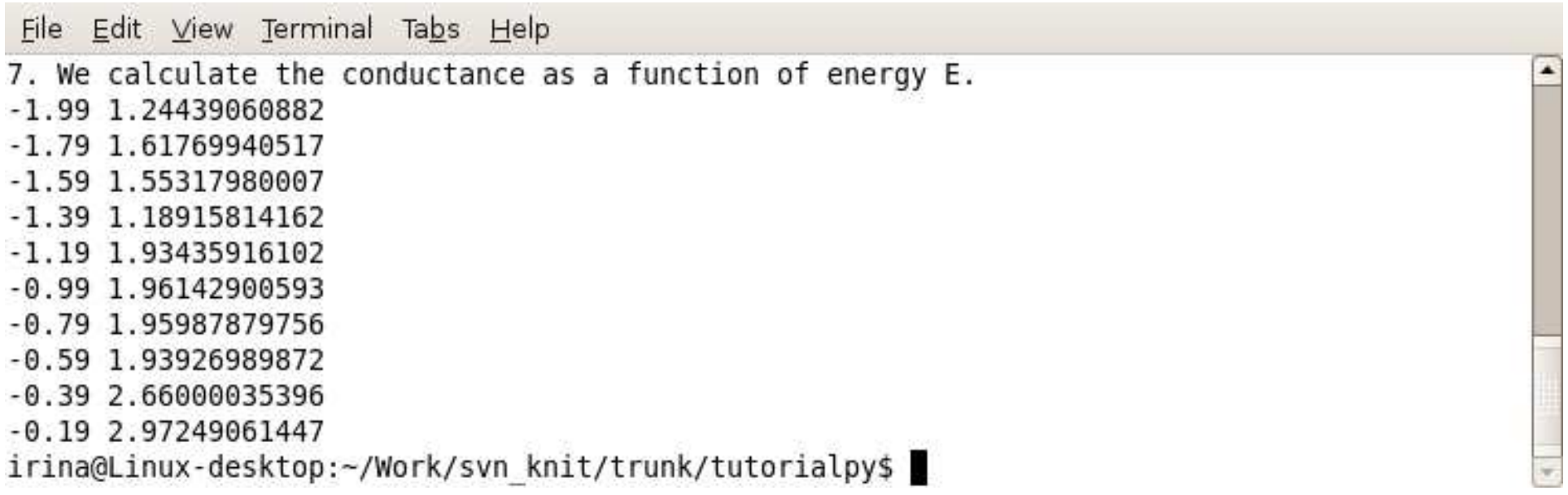

```
File  Edit  View  Terminal  Tabs  Help
7. We calculate the conductance as a function of energy E.
-1.99 1.24439060882
-1.79 1.61769940517
-1.59 1.55317980007
-1.39 1.18915814162
-1.19 1.93435916102
-0.99 1.96142900593
-0.79 1.95987879756
-0.59 1.93926989872
-0.39 2.66000035396
-0.19 2.97249061447
irina@Linux-desktop:~/Work/svn_knit/trunk/tutorialpy$
```

**FIG.3:** Conductance matrix for MyA

This example above illustrates the calculation of a conductance matrix for the system, defined in the 2-dimmentional space. Apart from conductance matrix, other transport properties and observables of systems can be calculated with KNIT as illustrated in Fig. 1.

## *2.3 A flavor of the Theoretical framework*

The system of N sites is described by the tight-biding Hamiltonian with nearest neighbor's interaction of the following form:

$$H = \sum_{<ij>}\sum_{\sigma\eta} c_i^{\sigma +} t_{ij}^{\sigma\eta} c_j^{\eta} + \sum_{i}\sum_{\alpha\beta} c_i^{\sigma +} V_i^{\sigma\eta} c_i^{\eta} , \qquad (1)$$

For any site with index *i* an operator $c_i^+ = \underbrace{(..,c_i^{\sigma+},..)}_{m}$ is a vector composed of $c_i^{\sigma+}$ operators. **m** defines the number of internal ***degrees of freedom*** at a site of the system[1] (for example: spin, e-h, s,p,d, etc). $c_i^{\sigma+}$ creates a particle with internal degree of freedom $\sigma = 1..m$ on ***i***-th site.

$t_{ij}$ is a **m x m** hopping matrix ;

$V_i$ is a **m x m** on-site energy matrix.

The system is connected to environment via semi-infinite leads. The leads can be analytically "integrated" out and appear in the formalism as self-energies $\Sigma_l(E)$, which are functions of energy. That provides boundary condition for the interface sites (sites which are common for the system and for leads). KNIT calculates the following retarded Green function of the system connected to external leads:

$$G(E) = (E - H + i\sum_l \Sigma_l(E))^{-1} \qquad (2)$$

From now on we consider the transport at fixed energy ***E*** and omit it as an argument of Greens functions and self-energies. The Green function ***G*** is $(N \times m) \times (N \times m)$ matrix where $N$ is total number of sites in the system; **m** is the number of internal degrees of freedom. An element of Green function $G_{ij}^{\sigma\eta}$ correspond to the quantum mechanical amplitude to go from ***i***-th site in internal state $\sigma$ to *j*-th site in state $\eta$. KNIT allows calculating two types of elements of G-matrix:

(1) $G_{a_k,b_l}^{\sigma\eta}$, where $a_k$ and $b_l$ sites that belong to the interface between a system and a leads ***k*** and ***l***, $\sigma, \eta = 1..m$ .

(2) $G_{i,b_l}^{\sigma\eta}$, where $b_l$ is a site that belongs to the interface between a system and a lead ***l***; ***i*** is an arbitrary site inside the system; $\sigma, \eta = 1..m$ .

Self-energy of different leads is also calculated in KNIT:

[1]The KNIT examples considered in this documentation corresponds to m=1,2 . Using the code for larger values of m is straightforward but requires modifying the file swig_knit.i and recompiling.

(3) $\Sigma_{a_l b_l}^{\sigma\eta}$ , defines self energy matrix between different sites $a_l$ and $b_l$ belonging to the same lead ***l*** with internal degrees of freedom $\sigma, \eta = 1..m$ respectively.

Knowing (1) (2) and (3) one can calculate different local and global properties of a system. Here are several examples:

The transmission probability between different leads ***k,l*** can be calculated using (1) and (2):

$$T_{kl} = G_{kl}\Gamma_l G^+_{lk}\Gamma_k = \sum_{a_l b_l c_k d_k} \sum_{\alpha\beta\gamma\delta=0}^{m} G^{\gamma\alpha}_{c_k a_l}\Gamma^{\alpha\beta}_{a_l b_l}(G^{\gamma\beta}_{c_k b_l})^*\Gamma^{\delta\gamma}_{d_k c_k}$$

where $\Gamma^{\alpha\beta}_{a_l b_l} = \mathrm{Im}(\Sigma^{\alpha\beta}_{a_l b_l})$. The related conductance is given by Landauer formula:

$$g_{kl} = \frac{e^2}{h}T_{kl}$$

The non-equilibrium local particle density and local current densities are defined through non-equilibrium Green function. The non-equilibrium Green function $G^<$ can be calculated between arbitrary sites ***i*** and ***j*** inside the sample using (1) and (3). The general formula using matrix multiplications reads as:

$$G^< = \sum_l G\Sigma^<_l G^+ = \sum_l f_l(iG\Gamma_l G^+) = \sum_l f_l G^<_l$$

with $\Gamma_l = \mathrm{Im}(\Sigma_l)$ and $f_l = \frac{1}{1+e^{(E-\mu_l)/kT_l}}$ is a distribution function in *l*-th lead that is characterized by chemical potential $\mu_l$ and temperature $T_l$.

An element of $G_l^<$ can be calculated as follows:

$$G_{l\ ij}^{<\alpha\beta} = i(G\Gamma_l G^+)_{ij}^{\alpha\beta} = i\sum_{a_l b_l}\sum_{\gamma\delta=0}^{m} G_{ia_l}^{\alpha\gamma}\,\mathrm{Im}(\Sigma_{a_l b_l}^{\gamma\delta})(G_{jb_l}^{\beta\delta})^* .$$

**NOTE: when $i$ appears in a regular expressions (not as an index), it denotes imaginary unit.**

# 3. KNIT distribution

Once you have downloaded and untared the KNIT archive (typically with the command **tar –xzf knit.tar.gz** under linux) you will find a tree of directories. Below is a brief description of their contents.
The directory wich ends with **py** contains python scripts. Those are the interesting ones for most users. The other directories are mostly used by the compiler.

| Folders/Files: | Current version of knit |
|---|---|
| documentation | Contains this document as well as all the python examples explained in this document. It also contains other documentation material like installation instructions; comments on KNIT functions; modifications done to KNIT during development (knit_README) |
| interface | Contains scripts files for the automatic wrapping of the C++ core into python. This directory is only used at compilation times. Advanced users might want to modify the file **swig_knit.i** which defines the links between all C++ objects (classes, functions and methods) and their corresponding Python objects, generated by swig. (This is necessary to compile the codes with more than two orbitals per site) |
| kernel | Contains all main C++ files – KNIT core functionality. |
| lib | Contains some general C++ files, not specific to KNIT. |
| libpy | Contains KNIT Python module. Once the compilation has been done, libpy is the only thing that remains useful for the user (you can copy it somewhere else and use knit functionalities from there). In particular, the file **knit.py** contains the main module, obtained by the automatic swig wrapping. |
| examples | C++ examples of KNIT for developers, who prefers not to use Python but C++. Those were used in the development stage of KNIT. Mostly obsolete now. |
| examplespy | Contains examples of systems modelled with KNIT (advanced examples with limited or no comments). Whenever you want to create a new system, you can have a look at this directory and see if part you need has not been done already. Includes examples with graphene, magnetic field, non collinear spin structures, superconductivity, basic parallelism... |
| tutorialpy | Contains well-commented basic examples helpful for new users. This was the only documentation before this document was written. A quick way to learn KNIT is to study and run these examples. |
| *knit_makefile* | This file contains the Makefile of KNIT. It should be modified before compilation to tell the compile where to find the various needed libraries– see Section 3. |
| *Makefile* | This file is almost empty and links to knit_makefile |

# 4. System requirements and installation guide

In this section we explain how to set up KNIT on your computer and make it ready to work. The KNIT system is usually easy to install on Linux computers (it takes about 20 minutes to install the various needed libraries on, say Ubuntu). If you don't know much about compiling, we recommend that you stick to Linux. KNIT has been known to work on Mac OS and we provide a few guidelines for that. As for Windows, well it is probably possible to compile KNIT for windows, but nobody really bothered. We did compile KNIT on Windows using cygwin (which emulates a linux API on windows) once, but the result is particularly slow. We provide a few guidelines for that as well but don't recommend it.

Once you have installed all the third party libraries, you need to edit the file knit_Makefile and modified a few lines in the beginning (to tell which version of the libraries you have and where you have installed them). Once this is done you should type **make python** in your terminal and you should be done.

Section 4.1 specifies a list of packages and libraries that are required by KNIT system. In section 4.2 we describe the setup process step by step.

## *4.1 System requirements*

Table 1 presents the list of required packages and the recommended order of their installation.

**Table 1.**

| | Package / library name | Short description | Installation comments |
|---|---|---|---|
| 1 | C++ compiler (cpp, gcc 4.*) | Needed to compile the kernel KNIT library | • usually a part of standard LINUX distributions |
| 2 | Python v. 2.5 or later. (Not the version 3) | Needed to build the application | • usually a part of standard LINUX distributions;<br>• to check version: python –V<br>• http://www.python.org/download/ |
| 3 | Python-dev | that includes header Python.h | Available in standard installers as ***apt-get*** |
| 4 | Python – Numeric, numpy, | Python library for numerical computing | • Available in standard installers as ***apt-get***<br>• should be installed in its default path |
| 5 | Blitz++ v.0.9 or later | Template library for numerical computing. Contains matrices and tensors. | http://www.oonumerics.org/blitz/download/ |
| 6 | PyX-0.9 | PyX is a Python package for making 2d and 3d plots (postscript and PDF | • http://pyx.sourceforge.net/<br>• if Pyx-10 is used change color.palette to color.gradient |
| | | files). | • PyX is not completely necessary but it is nice ans some examples won't work without it. |
| 7 | lapack 2.5 | Linear Algebra PACKage | • http://www.netlib.org/lapack/<br>• sometimes one needs to rename or create an appropriate symbolic link to liblapack.so.n => liblapack.so , where -n- is the version number<br>• This should also install the important BLAS on which lapack relies heavily |
| 8 | libg2c (mostly obsolete know, better use *gfortran*) | Required for linking C with lapack | • sometimes one needs to rename or create an appropriate symbolic link to libg2c.so.n => libg2c.so , where -n- is the version number |
| 9 | gfortran | Required for linking C with lapack | • Available in standard installers as ***apt-get*** |
| 10 | Swig 1.3.39 | *(Simplified Wrapper and Interface Generator)* Required for C++ - Python code binding | http://www.swig.org/ |

Why do I have to install all THIS? Well if you want a pure C++ version of KNIT, you need only blitz++ (we used it for the tensor template class) and Blas/lapack (which is THE standard library for linear algebra). The program swig is used to generate a python wrapping of the C++ code (i.e. python class that mimick the C++ class. Those python classes do nothing except call the correct C++ function when needed). Section 8 of this document contains a little more on KNIT architecture.

### *4.2 Installation*

# A. LINUX

**Example: Step by step setting up KNIT for LINUX – Ubuntu**

1. **Preset the environment:**

- Check for ***gcc***;
- Check for ***Python*** (python -V) – if not on the machine – install Python 2.5 (or higher).
- Install other Python – ***dev, Numeric, numpy*** (from the standard installer as apt-get)
- Install ***Blitz++, PyX, lapack***, and ***swig*** following the instructions provided by these packages;

- Check for ***gfortran*** library.

Sometimes one needs to create an appropriate symbolic links :

```
liblapack.so.XXX => liblapack.so
libg2c.so.XXX => libg2c.so ,
libgfortran.so.XXX => libgfortran.so
```

Here is an example:

```
user@Linux-desktop:~/Work/svn_knit/trunk$ sudo ln -sf /usr/lib/libgfortran.so.2
/usr/lib/libgfortran.so
```

In our example, the package is extracted into **`~/Work/svn_knit`** folder.

## 2. Changes in KNIT package:

- Extract the svn_knit package in some directory.
- In */svn_knit/Trunk modify ***knit_Makefile*** by customizing it for your particular machine and OS distribution:

Example:

```
# ***********************************************************
#  1. MACHINE SPECIFIC PART:                               *
#  ADD LIES CORRESPONDING TO YOUR MACHINE HERE             *
# ***********************************************************
# define HOST variable in order not to use all the time $$HOSTNAME
HOST = $(shell echo $$HOSTNAME)

ifeq ($(strip $(HOST)),)
$(error HOSTNAME is not defined !)
endif

# DIRECTIVES FOR MY USER LINUX-DESKTOP
ifeq ($(strip $(HOST)),Linux-desktop)
CC= g++
PYTHDR=/usr/include/python2.5/
BLZHDR=/usr/include/
BLZLIB=/usr/lib/
MYBLAS=/usr/lib/
LFLAGS = -L$(HOME) --export-dynamic  -L$(MYBLAS) -llapack -lgfortran -
lpython2.5 -L$(BLZLIB) -lblitz
endif
```

Explanations:

`HOST` – name of the computer. Note, that on different Linux distributions command:

`echo $HOSTNAME`

may work differently. For example on Ubuntu Linux instead of :

`echo $HOSTNAME`

one should use:

hostname
In the example when `HOST` is equal to `Linux-desktop` the following options are used:
`CC` – name of c++ compiler
`PYTHDR` – location of `Python.h` (to find it use `locate Python.h`)
`BLZHDR` – location of `blitz/blitz.h`
`BLZLIB` – location of `libblitz.so`
`MYBLAS` – location of `libblas.so` and `liblapack.so`
Other libraries such as `libpython2.5.so` `libgfortran.so` are supposed to be location described by your `$PATH` system variable.

Some debugging options can be also selected here:

```
# ************************************************************
#  2. CHOOSE YOUR OPTIONS HERE                               *
# ************************************************************

CFLAGS = -I$(PYTHDR) -I$(BLZHDR) -I$(CURDIR) -I$(CURDIR)/kernel -
I$(CURDIR)/lib
#CFLAGS +=  -O2 -fPIC  -pg
CFLAGS +=  -O2 -fPIC
#CFLAGS +=  -O3 -fPIC -finline-functions
#CFLAGS +=  -g -fno-inline -fno-default-inline -fPIC
#CFLAGS += -D_DEBUG_DI_ -D_DEBUG_LONG_ -DUSE_BLITZ  -O2 -fPIC
CFLAGS += -D_DEBUG_
#CFLAGS +=  -pg
```

However we do not think that this part of make should be changed.

## 3. Compile and build KNIT:

We compile the application using **`make clean`** and **`make python`** commands
Example of the trace:

```
user@Linux-desktop:~/Work/svn_knit/trunk$ make clean
…
user@Linux-desktop:~/Work/svn_knit/trunk$ make python
…
```

## 4. Test installation:

- Test if the application was built correctly. To do so, we propose to start Python and try to import functions from libpy library:

```
user@Linux-desktop:~/Work/svn_knit/trunk$ python
Python 2.5.2 (r252:60911, Jul 31 2008, 17:28:52)
[GCC 4.2.3 (Ubuntu 4.2.3-2ubuntu7)] on linux2
Type "help", "copyright", "credits" or "license" for more information.
>>> from lippy import *
Traceback (most recent call last):
  File "<stdin>", line 1, in <module>
ImportError: No module named lippy
>>> from libpy import *
>>>
```

- If the test went fine, we execute the example:

```
user@Linux-desktop:~/Work/svn_knit/trunk$ cd examplespy/
user@Linux-desktop:~/Work/svn_knit/trunk/examplespy$ python benchmark_bar.py
 SYSTEME OF SIZE 100 x 100
 CONSTRUCTION = 0.856734037399
 CONDUCTANCE CALCULATION = 63.9291319847
Comment for the history file?
 (type <enter> not to add the data)slow Linux-desktop
user@Linux-desktop:~/Work/svn_knit/trunk/examplespy$ emacs benchmark_bar.
benchmark_bar.history  benchmark_bar.py
user@Linux-desktop:~/Work/svn_knit/trunk/examplespy$ emacs benchmark_bar.history &
[1] 5909
user@Linux-desktop:~/Work/svn_knit/trunk/examplespy$ gedit benchmark_bar.history &
[2] 5910
[1]   Done                    emacs benchmark_bar.history
user@Linux-desktop:~/Work/svn_knit/trunk/examplespy$
```

Some useful examples can be also found in `~/Work/svn_knit/trunk/tutorialpy$`. We will consider these examples and explain how to create your own KNITable files in the next sections. However, it is also a good place to finish our installation process by executing the following:

```
user@Linux-desktop:~/Work/svn_knit/trunk$ cd tutorialpy/
user@Linux-desktop:~/Work/svn_knit/trunk/tutorialpy$ python simplest_example1.py
```

# B. MacOS

This is a short help to compile knit on MacOS

## 1. Preset the environment:

- Install Xcode from the system DVD
- Check that ***libblas.dylib*** and ***liplapack.dylib*** are installed in ***/usr/lib*** (Optionally: an optimized library ***libmkl_lapack.dylib*** can be installed by installing "Intell C++ compiler for Mac" from the intell web site. A commercial version only; but exists a trial version.)
- Using ***macports*** install:
    a. python: `sudo port install python`
    b. Blitz: `sudo port install blitz`
    c. swig: `sudo port install swig`

- Install ***numpy*** (an old Python package, ***Numeric*** , will not work)
- Install PyX
- Check the paths for
  Python.h
  Numeric/arrayobject.h
  blitz/array.h

## 2. Changes in KNIT package:

- Extract the svn_knit package in some directory.

- file: interface/knit_Python_Data.C
  In the header add new line: `typedef unsigned int uint;`

- file: interface/knit_system.h
  Change:
  `static int const not_lead=-1;`
  to
  `static char const not_lead=-1;`

- In */svn_knit/Trunk modify ***knit_Makefile*** by customizing it for your particular machine and OS distribution:

Example:

```
# *********************************************************
#  1. MACHINE SPECIFIC PART:                              *
#  ADD LIES CORRESPONDING TO YOUR MACHINE HERE            *
# *********************************************************
#------------------
# DIRECTIVES FOR MACBOOK
ifeq ($(strip $(HOST)),drecam0063.saclay.cea.fr)
CC= g++
PYTHDR=/Library/Frameworks/Python.framework/Versions/2.5/include/python2.5
# Numeric package must be installed here. Python.h is here;
# Numeric/arrayobject.h is here

BLZHDR=/opt/local/var/macports/software/blitz/0.9_0/opt/local/include
# blitz/array.h  is here

BLZLIB=/opt/local/var/macports/software/blitz/0.9_0/opt/local/lib

# libblitz.dylib is here.

# Use -lgfortran instead of obsolete -lg2c
LFLAGS = -L$(HOME)  --export-dynamic  -framework Accelerate -llapack -lblas  -
lgfortran -lpython2.5   -L$(BLZLIB) -lblitz -dynamiclib -flat_namespace
endif
```

## 3. Compile and build KNIT:

```
make python
```

## 4. Test the installation:

```
cd examplespy/
```

```
python benchmark_bar.py
```

**NOTE:** Under MacOS KNIT works significantly slower than under LINUX.

# C. Windows

This installation might be unstable and even if made successfully, **it is extremely slow**. Though, we strongly believe that none will read this section.

This is a short help to compile knit on windows using cygwin. Note: the result only work on cygwin python, not on win32 python. One would need a different build to do so (possibly with visual C++). If you are used to comile programs in Windows, you can probably do a much better job than this.

## 1. Preset the environment:

- Installing Blitz:

Straightforward: download then type

```
./configure CXX=g++
make install
```

I had one problem: the configure script did not work because of DOS ending instead of UNIX ending. (whatever that means)

This can be cured with d2u:

```
d2u configure
d2u config/*
```

- SWIG, BLAS, LAPACK already available in cygwin.
- Install PYX : download and type

```
python setup.py install
```

## 2. Changes in KNIT package:

- Extract the svn_knit package in some directory.
- In *\svn_knit\Trunk modify ***knit_Makefile*** by customizing it for your particular machine and OS distribution:

Example:

```
# ********************************************************
#  1. MACHINE SPECIFIC PART:                             *
#  ADD LIES CORRESPONDING TO YOUR MACHINE HERE           *
# ********************************************************

# DIRECTIVES FOR MY LAPTOP (INTEL DUAL CORE 2 GHz)
ifeq ($(strip $(HOST)),DREC-HB-002386)
CC= g++
PYTHDR=/usr/include/python2.4/
PYTLIB=/bin/
BLZHDR=/usr/local/include/
BLZLIB=/usr/local/lib
LFLAGS = -L$(HOME) --export-dynamic  -llapack  -lblas -lg2c -L$(PYTLIB) -
lpython2.4 -L$(BLZLIB) -lblitz
endif
```

I found the location of libpython.so with

```
find / -name "libpython*"
```

### 3. Compile and build KNIT:

```
make python
```

- Change the name of `_knit.so` to `_knit.dll`

### 4. Test the installation:

```
python benchmark_bar.py
```

# 5 Getting started with modeling

## *5.1 C++ kernel vs. Python interface*

The **kernel** of the KNIT modeling system is written in **C++** and represents the main functionality of KNIT. A user accesses this functionality via an **interface**, written in **Python**. Technically, user has to write a ***simple script*** – a Python program – that will be interpreted by KNIT. KNIT passes all Python function calls from the user script to its kernel and returns the results in a form defined by the user (e.g. graphs, plots, flat data files). Therefore, all the implementation details are hidden from the user and a knowledge of C++ is not required. In more details, the KNIT architecture is addressed in section 8 of this document.

## *5.2 Five modeling steps*

KNIT calculates local and global transport properties of a mesoscopic quantum systems. In KNIT, one considers a quantum ***system*** of N ***sites*** connected to several conducting ***electrodes***, also called ***leads***. User specifies the system (its geometry and leads) in a ***script file***, written in the ***Python*** language.

To write a Python script for KNIT, user has to accomplish the following steps:

A. System construction
B. Lead construction
C. Total system construction
D. System solving
E. Visualisation of results

In a tutorial below we illustrate how to proceed with these modeling steps. We divide this tutorial into 5 parts – by the number of steps to accomplish. Each part contains a theory and a Python code related to a modelling step.

## *5.3 Tutorial: Calculation of conductance matrix for a 2D system with two leads*

## A. SYSTEM CONSTRUCTION

### System and Site

In KNIT, a system can be specified in 2D, 3D, … space. In 2D space, for example, the system is characterized by its height and with (W, H) and represents a finite number of sites, enumerated from 1 to N = W x H:

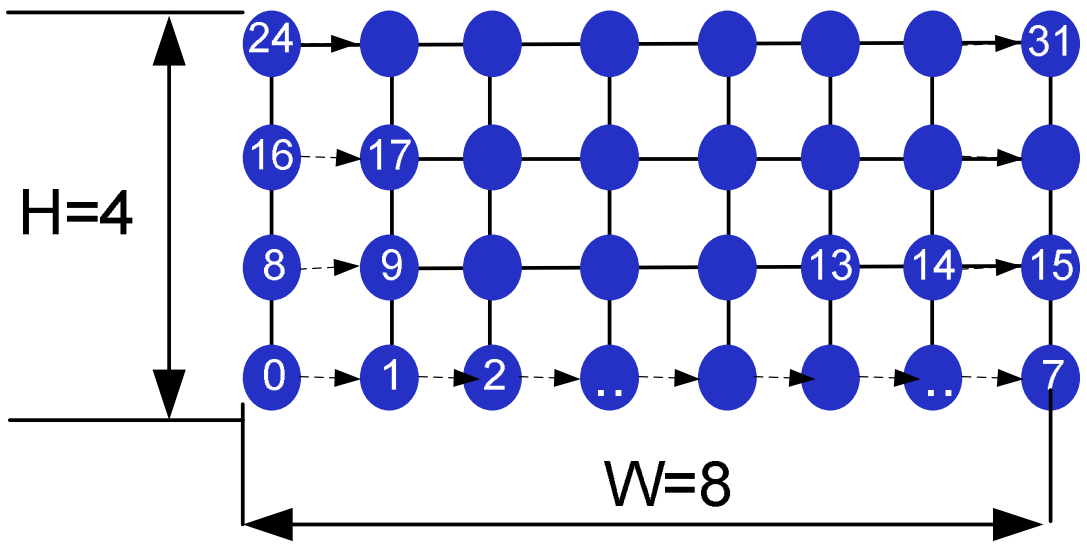

**FIG.4:** A simple example of a 2D - system A specified as a rectangle 8x4 with 32 sites.

Along these lines, the system in 3D space is specified by a triple (H, W, D) and the set of W x H x D sites. All the reasoning below can be extended for the 3D systems.

Each site i, **i=1..N-1** of the system in KNIT is specified using a ***micromatrix*** $\mathbf{V_i}$, and a vector $\mathbf{t_{ij}}$**,** j=0..maxN of hopping matrices.

Micromatrix $\mathbf{V_i}$ corresponds to the on-site energy matrix from the equation (1) for the site **i**; it has a dimension **m x m** and reflects the microscopic degree of freedom of a system as explained in section 1.3.

Vector $\mathbf{t_{ij}}$ , **j=0..maxN** contains hopping matrices (Eq.(1)) between the site i and its j-th neighbour . For a given site, each element of $\mathbf{t_{ij}}$ specifies probability amplitudes for an electron to hop from this site to its neighbor j. **maxN** is a maximum number of neighbours for a given geometry: for regular 2D-space maxN = 4; for regular 3D-space maxN=6; for graphene hexagon structure maxN = 3.

**Example:** To construct a rectangular system, we should specify the ***sites*** by defining intersite hopping **t**, and onsite energy **V** and define the ***geometry*** of the system.

For each site, we should specify its hopping matrix **t** and its on-site energy matrix **V:**

We specify this system as constructed of three blocks A, B, and C in 2d-space. To specify our sites, we consider d=1.

```
t=knit.scalarM(1.)
V=knit.scalarM(0.)
```

**NOTE:** Dimension **m** cannot be changed. Both t and V should be either **scalarM** or **vectorM.**

Values of t and V are matrices **m x m,** where the selected dimension **m** defines a number of internal degrees of freedom at the site of the system (for example: spin, e-h, s,p,d, etc). For the moment, KNIT supports d=1,2 but can be generalized for arbitrary value of d.

In Python specification, **m=1** corresponds to the <site type> = **scalarM.** In this case values of t and V are scalars. To specify **d=2,** we use <site type> = **vectorM.** In this case values of t and V are matrices 2 x 2.

## Building a system

System construction is a process that connects sites together by assigning the list of neighbors to each site. During construction each site receives its number. Enumeration is continuous (Fig. 4).

**<u>Example</u>**: We define the geometry of a system – its width **W**, height **H**,(and depth **D** for the 3D space) - as scalar values that correspond to the number of sites on a given side of the rectangle, for example:
In general, to build a system, we use function **knit.rectangle** with the following signature:

knit.rectangle([W, H, D],[x0,y0,z0],t,V)

Here x0, y0,z0 – are coordinates of the bottom left corner of our rectangle.

In our example, we parameterize the system by using a size unit **MYSIZE**:

```
MYSIZE=1
```

This is advantageous if one wants to scale the system, i.e. Increase the number of sites by keeping the same relative system geometry.
We build three blocks to produce a resulting geometry as illustrated in fig. 5-a:

```
MyA=knit.rectangle([MYSIZE *8,4* MYSIZE],[0,3* MYSIZE],t,V)
MyB=knit.rectangle([3* MYSIZE,7* MYSIZE],[5* MYSIZE,0],t,V)
MyC=knit.rectangle([2* MYSIZE,6* MYSIZE],[0,3* MYSIZE],t,V)
```

## Building a system of required geometry

In general, the system of a required geometry can be constructed by:

- Gluing one block on top of another;
- Adding sites;
- Moving sites;
- Moving all system - shifting

## Building a system of required geometry – coller:

***Gluing*** is a process of connecting 2 or more system blocks together to produce a system with a required geometry. Fig. 5 illustrates a system built out of two blocks A, B. Blocks are glued one by one: in our example in Fig. 6, the block B is glued on top of A. In a resulting system, the overlapping parts between A and B are replaced with those of B: this literally means ***gluing on top***.

Technically, gluing the lock B on top of the block A from the example in Fig. 6 can be seen as 3 tasks:

1. all sites of A that are overlapping in space with sites of B will be replaced with those of B ;

Knitting of the AB system:

2. All sites of the resulting system will be re-enumerated to avoid site number duplication;
3. Neighbors for each site will be reassigned.

**Note:** the enumeration of sites of the final system in KNIT depends on the order in which the rectangles are glued and should be verified in each case.

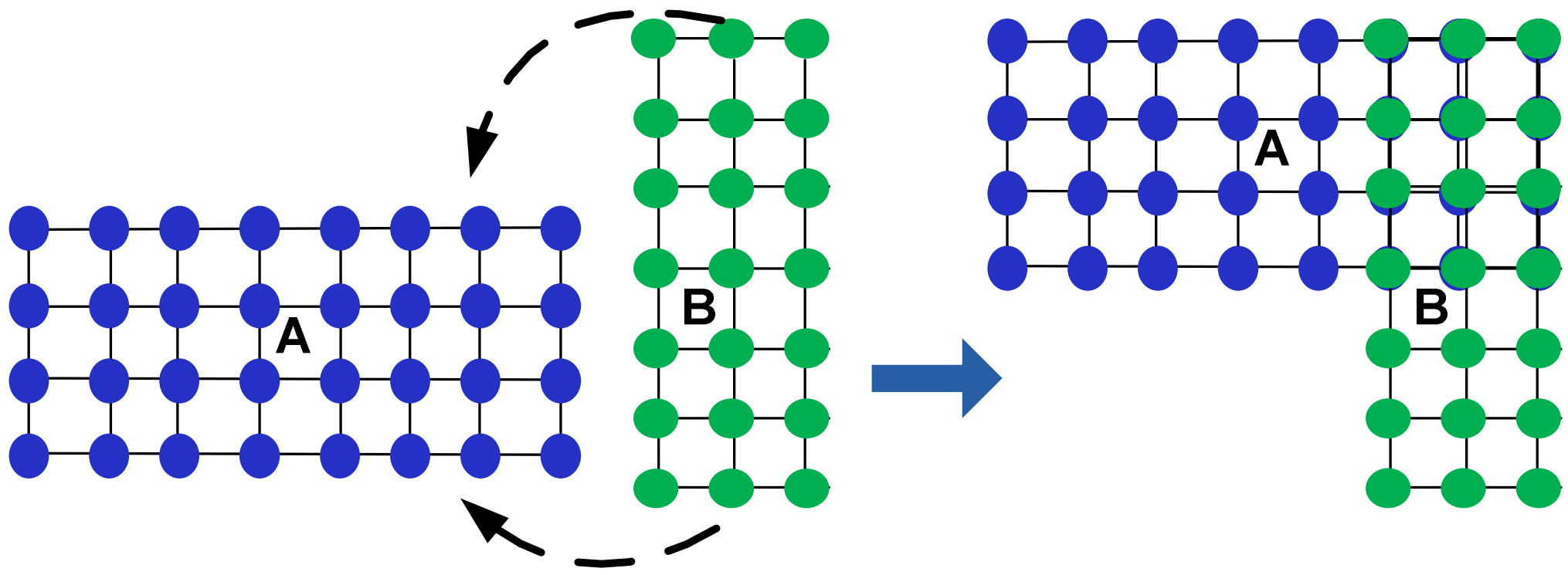

**FIG.5:** A System built out of 2 blocks: A and B by gluing.

To build a system of a required geometry, modeler has to define necessary blocks and position them in (2D or 3D) space by specifying physical coordinates of the bottom-left corner of each block $\mathbf{x_0,y_0}$ (or $\mathbf{x_0,y_0,z_0}$ for 3D)[2] . Based on these coordinates, the overlapping parts of the blocks are identified and gluing is made. ***If there is no overlapping between blocks – sites cannot be connected for future knitting.***

**<u>Example</u>**: We glue B and C on top of A as follows:

```
MyA.coller(MyB)
MyA.coller(MyC)
```

The resulting ABC system in fact is constructed on the base of the system myA; so we can write:

```
MyABC = MyA
```

[2] Technically, there is a one to one correspondence between the site number **i** and its physical coordinates **(x,y,z)** .

To verify that the resulting system has a required geometry, the system can be plotted using a python function ***visu2D***:

```
system.visu2D(MyABC, “<SOME FILE NAME>”)
```

visu2D makes a plot of our system MyABC and writes this plot into the files `<SOME FILE NAME>`***.eps*** and `<SOME FILE NAME>`***.pdf***

Based on the coordinates of the blocks A, B, C, specified above, we obtain the system configuration as illustrated in Fig. 6-a.

Assuming that the block A was positioned incorrectly:

```
MyA=knit.rectangle([MYSIZE *8,4* MYSIZE],[3,3* MYSIZE],t,V)
```

as a gluing result, we obtain the system configuration as shown in Fig. 6-b. This configuration is incorrect as block C cannot be glued to AB.

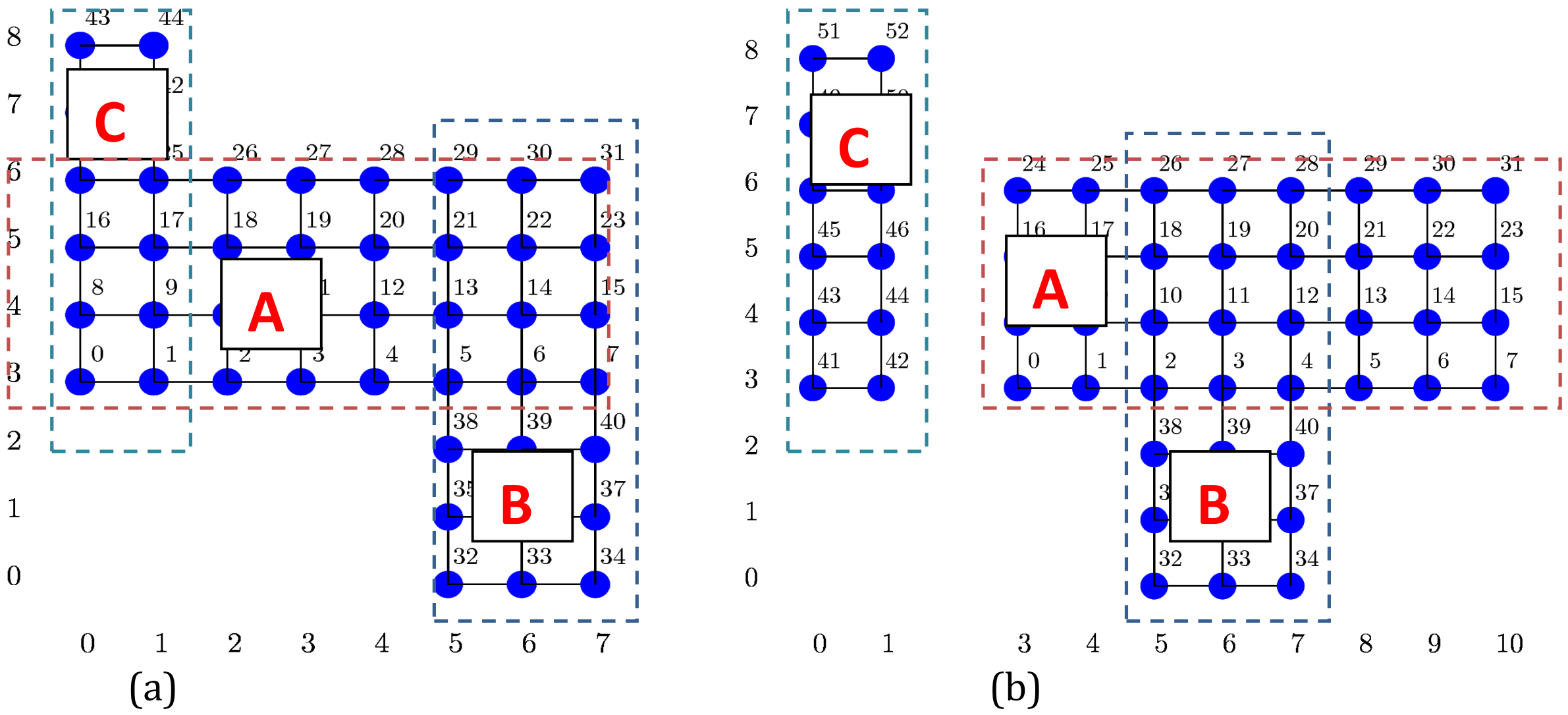

**FIG.6: Gluing of blocks B and C on top of the block A.** The images are produced by KNIT: the system sites are connected if they have an edge between them. Otherwise the sites are disconnected. a) system ABC is built correctly. b) Due to incorrect positioning of the block A, the block C is disconnected from the system. This system will not be solved!

**NOTE: geometry type (2D or 3D) must be the same for all the building blocks and the leads.**

## Building a system of required geometry – add_site:

***Adding a site***, we specify a new site in some position {newX, newY (, newZ – for 3D) }
This is illustrated in Fig. 7-b.

**Example**: We add a site to the system by specifying its coordinates and a neighbor list:

Coordinates

```
MyABC.add_site([4,38], t, V, [MYSIZE*4, MYSIZE*2])
```

Indices of neighbours

**Building a system of required geometry:**

***Moving a site*** technically means reassignment of its neighbors. In KNIT, we can move the site i to a new position {newX, newY (, newZ – for 3D) }
This is illustrated in Fig. 7-c.

**Example**: In the system (fig. 7- a) we move the site number 13 by assigning to it new coordinates and new neighbor list:

Index of the moved site

```
MyABC.move_site([27], t, V, [MYSIZE*3, MYSIZE*7], 13)
```

Indices of neighbours

Coordinates

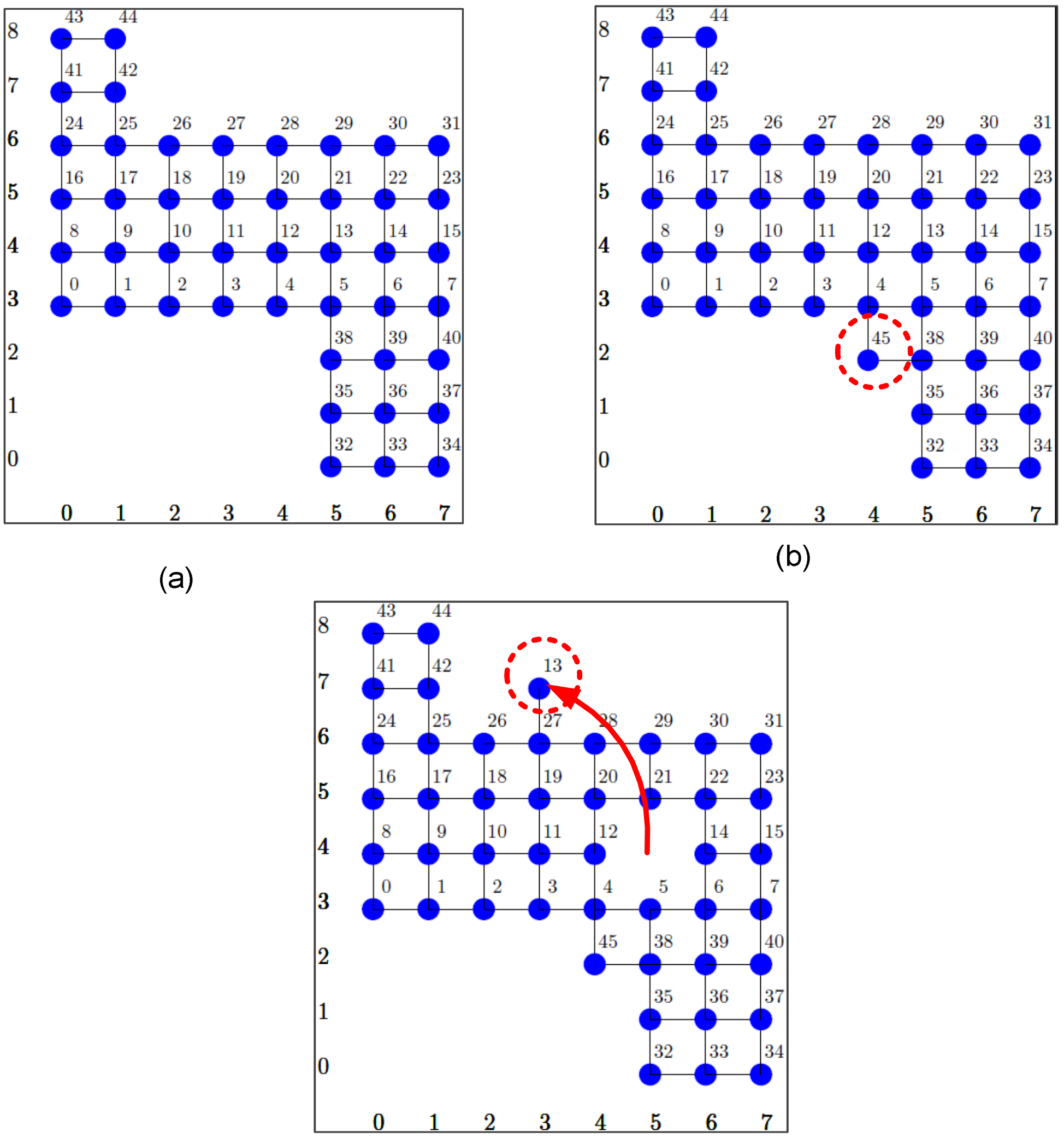

(a) (b) (c)

**FIG.7: System construction:** a) System is constructed by gluing rectangular blocks; b) One site is added c) Site number 13 is moved.

## Building a system of required geometry – shift_system:

***Shifting*** a system on the vector Δ (delta) is equivalent to moving all the system sites to this vector in coordinate space.

**Example:** We shift the system s below on the vector Δ = [2,1]:

```
s.shift_system([2,1])
```

delta

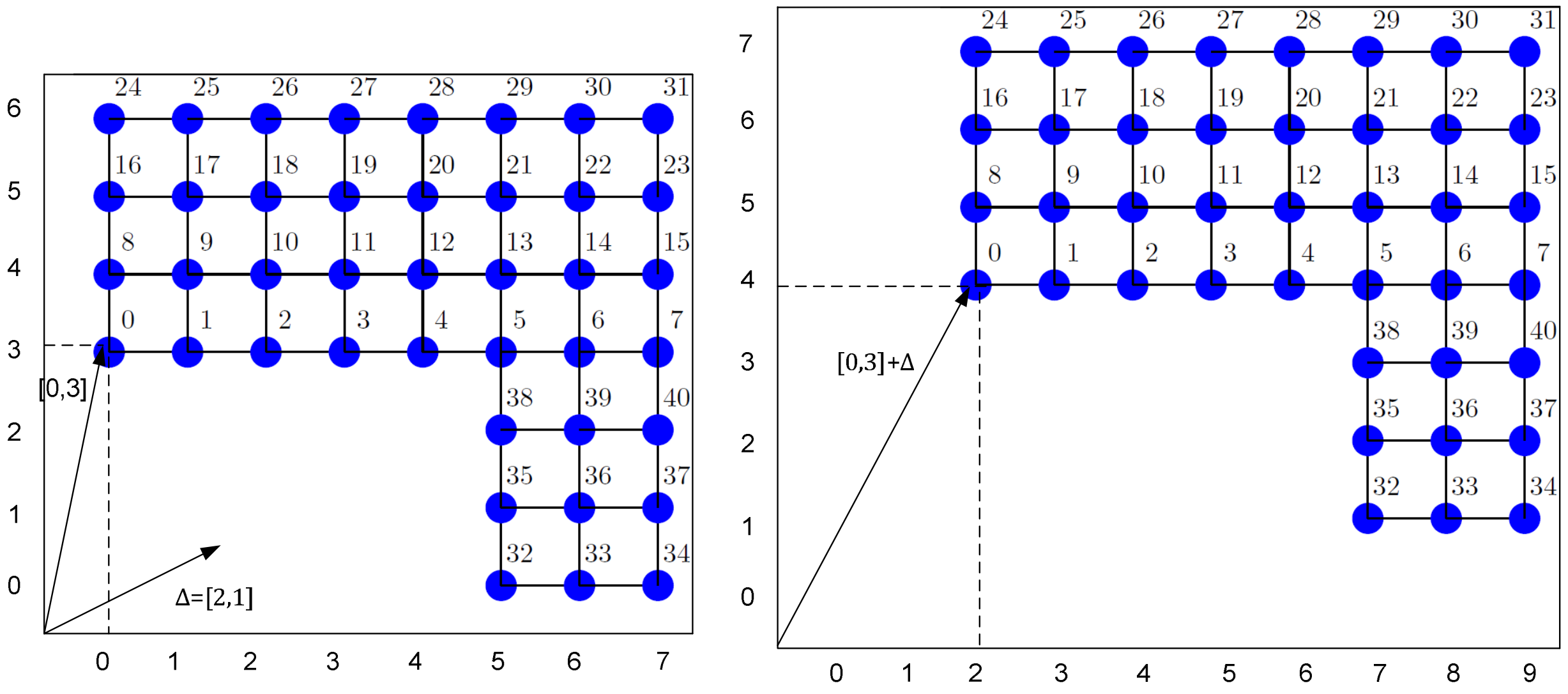

**FIG.7: System shift on** Δ = [2,1]

# B. LEAD CONSTRUCTION

A lead is a semi-infinite wave-guide that is attached to a mesoscopic system. Each system must be attached to at least one lead in order to be solved by KNIT.

To construct a lead, we first create a ***lead interface*** - the part that will be attached to one of the system sides. The lead interface is a system with only one layer of sites (i.e. one of the lead sizes W=1 or H=1 or D=1), then we extend this interface (by copying the layer of sites) to obtain a ***semi-infinite*** contact.

This interface technically represents a rectangular system with one of its sizes equal to 1:

```
Interface_Contact=knit.rectangle([iW, iH],[leadX0, leadY0],t,V)
```

Here: **iW = 1 or iH = 1**. A pair [leadX0, leadY0] specify a position where the lead will be finally attached to the system (I.e. typically it should correspond to a bottom-left corner of one of the blocks of the system).

After specifying the interface of the lead, we construct the lead itself, using a function

```
Lead=unreservoirN(Interface_Contact,t)
```

formally speaking it repeats the `Interface_Contact` infinite number of times**.** The hopping between different slices of the lead is `t`. In our example , we specify the interfaces and create the leads as illustrated in fig. 9-a:

```
interface_contact1=knit.rectangle([1,MYSIZE*6],[0,MYSIZE*3],t,V)
lead1=knit.unreservoirN(interface_contact1,t)

interface_contact2=knit.rectangle([MYSIZE*3,1],[5*MYSIZE,0],t,V)
```

```
lead2=knit.unreservoirN(interface_contact2,t)
```

**NOTE: for a 2D-system interface must be also specified as 2D- rectangles. The same is valid for the 3D-system.**

We use the same hopping matrix t for specifying the system and the lead. Repeating this step, modeler can specify an arbitrary number of leads.

**NOTE: The interface sites belong both to the system and to the lead. The hopping matrix elements from the lead to the system are exactly the same as those inside lead itself.**

## C. INFINITE SYSTEM CONSTRUCTION

Similarly to a simple rectangular system, the lead should be ***glued on one of the system sides* –** i.e. it should the lead interface contact (that was duplicated to create a lead) have correct physical coordinates.

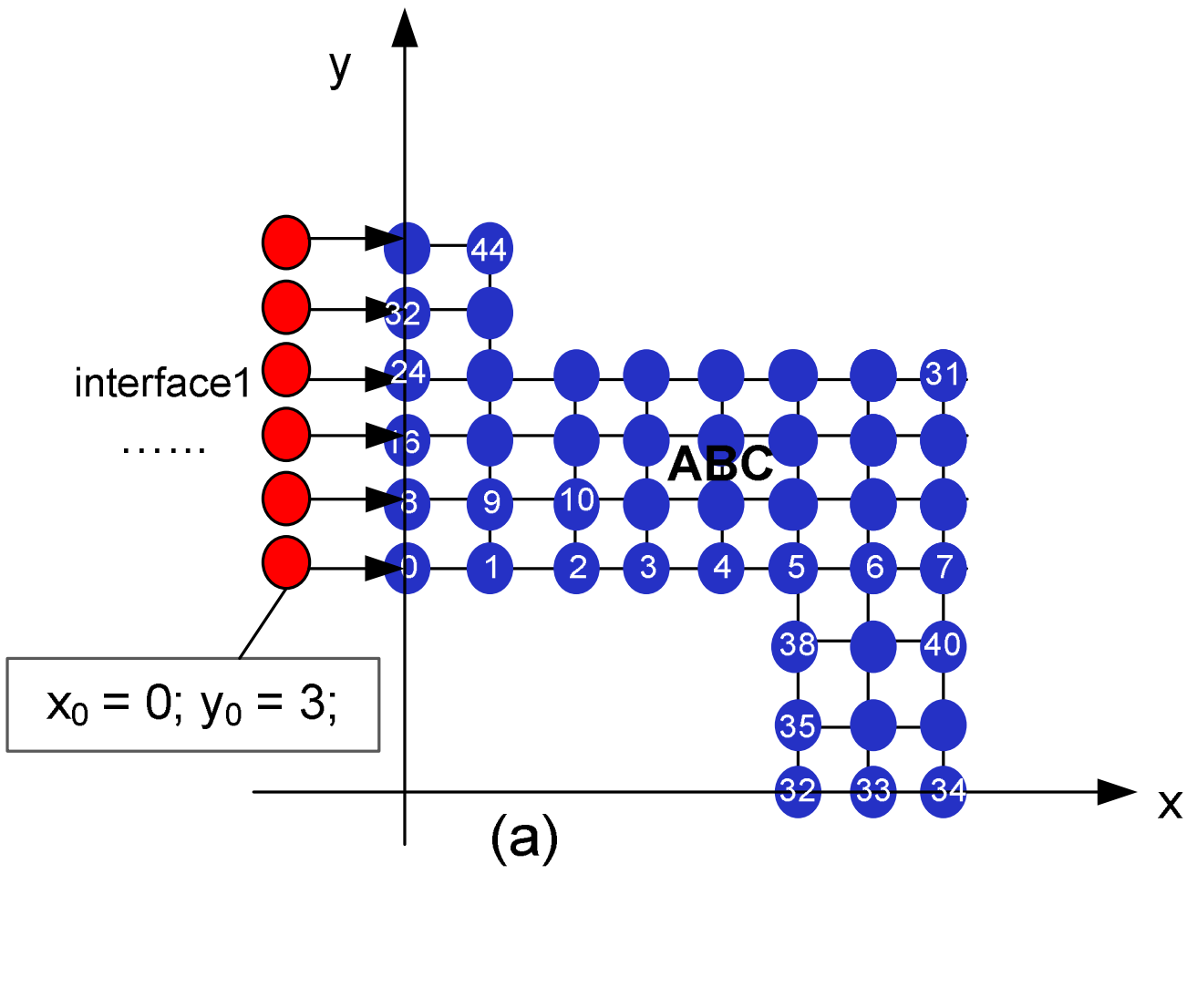

**FIG.8: Leads** (a) Specification of the first lead: it must overlap exactly one layer of sites of the system; (b) two leads are connected to the system. The lead sites substitute the system sites.

When the system is constructed, we attach the leads to our system and obtain the resulting system with leads:

Creation of infinite system with one lead

```
MyABCwithLeads=knit.systemtotal(MyABC,lead1)

MyABCwithLeads.addlead(lead2)
```

Adding the second lead

To add the second (and further) leads to the system, we use the method ***addlead***

**NOTE: in order to be solved, system must have a source of particles, i.e. at least one lead must be specified and attached to the system.**

To verify that the leads are correctly connected to the system, we can plot the system using ***visu2D***:

```
system.visu2D(MyABCwithLeads, “<ANOTHER FILE NAME>”)
```

**NOTE: for displaying 3D systems, <u>visu3D</u> should be used instead of visu2D. The signature is the same.**

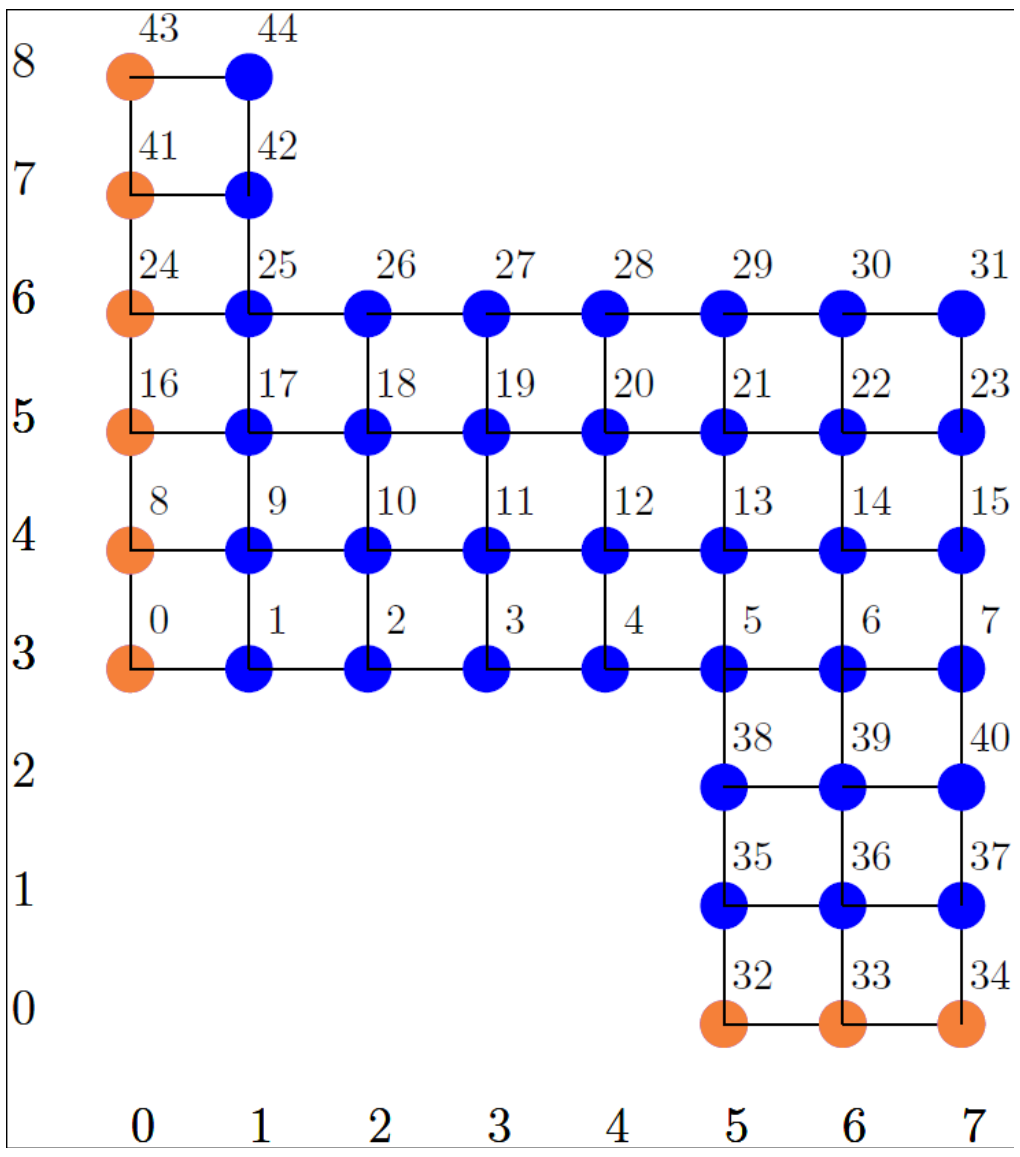

**FIG.9: System MyABCwithLeads**

Sites of interface contacts are shown here in orange color.

## D. SYSTEM SOLVING

KNIT solver calculates (i) global transport properties of the system using the ***knitting*** algorithm; (ii) local observables of the system using the ***sewing*** algorithm.

The knitting algorithm calculates the retarded Green function of the system connected to external leads as described in section 1.2. More precisely, the Green function $G_{a_l b_k}$ between site *a* of a lead *l* and a site *b* of a lead *k* is calculated (fig 9). Calculation is done by adding sites one by one to the "knitting thread". Green function $\overline{G_{a_l i}}$ for each added site *i* is calculated recursively, taking into account the calculations for all previously added sites. Assuming the knitting from *l* to *k*: at each knitting step, the algorithm momentarily stores and updates the value of Green function for all sites of the lead *l* and for those sites of the system, for which NOT all their neighbors are knitted yet (shaded circles). These sites are said to belong to the knitting ***interface.***

Knitting is continued until all the N sites are added. Eventually $G_{a_l b_k}$ is calculated. Those matrix elements give access to all transport properties like conductance or shot noise, but no information on what happens inside the sample. For these purposes, the ***sewing*** algorithm is implemented.

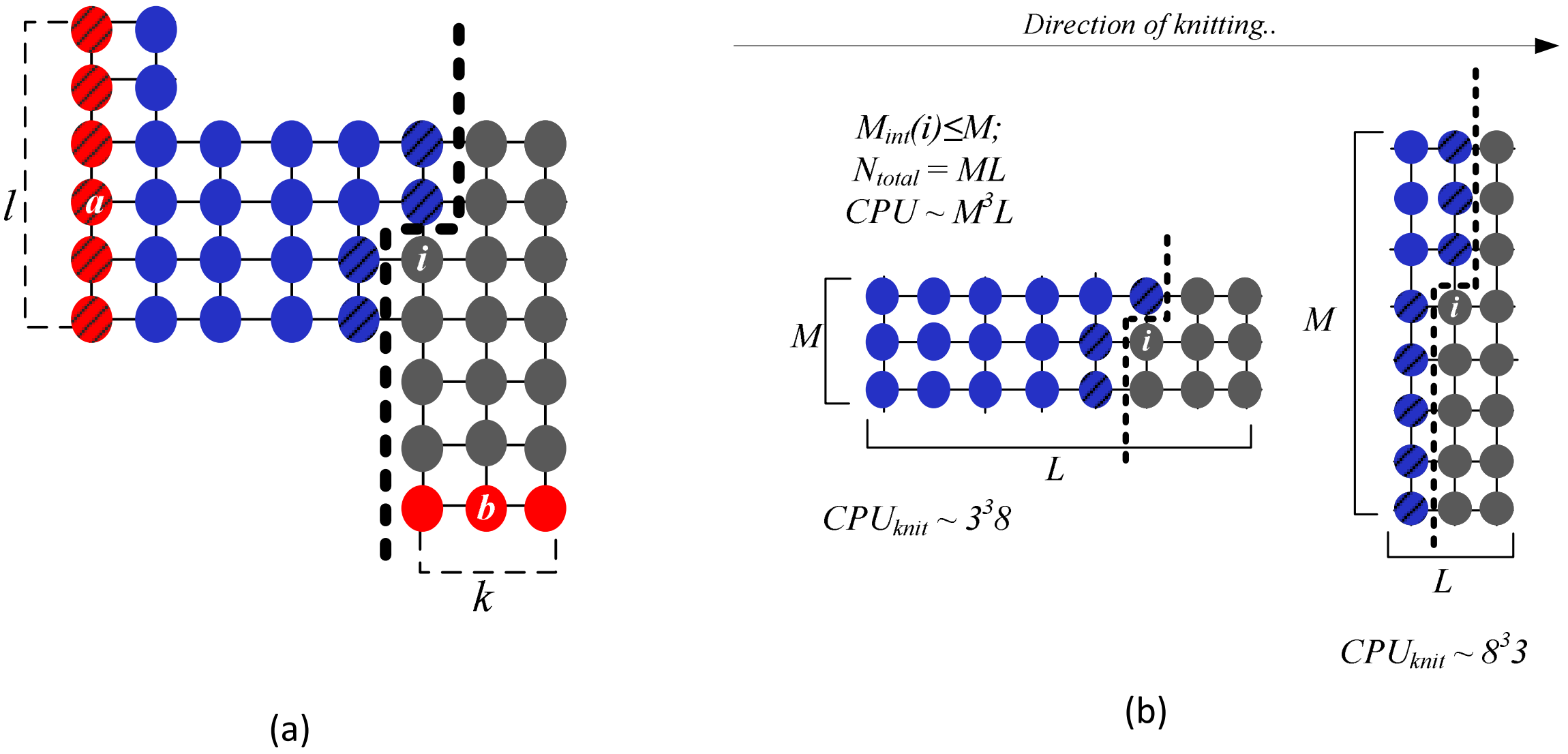

**FIG. 9:** (a)**Knitting.** System has two leads: *l* and *k*. The knitting algorithm adds sites one by one. The thick dashed line separates the part already included (left) from the part that is still to be knitted to the system (right). (b) Performance of knitting and sewing algorithms depending on the system geometry.

The sewing algorithm extends the knitting algorithm to the calculations of local observables. For a given site *i* of the system, the sewing algorithm stores a value of the partial Green function $\overline{G_{a_l i}}$ obtained by knitting. The second part of this algorithm calculates the exact value of the retarded Green function $G_{a_l i}$ by applying the knitting algorithm in the opposite direction: from the lead *k* to the lead *l*. Namely, we are "unknitting" the sites one by one starting from the site *b* of a lead *k* until we reach the site *i*. Using the sewing algorithm, the retarded Green function $G_{ij}$ between two arbitrary system sites *i* and *j* of the system can be calculated.

When the system and its leads are constructed, we proceed with system solving. In KNIT, knitting and suing routines can be accessed via Python interface: functions ***knitonly/knitonly_S*** and ***knitANDsew/knitANDsew_S*** (_S defines the corresponding function for systems with spin).

In our example, we calculate the conductance between leads (a matrix ***G_MATRIX***) as a function of energy **E**:

```
for i in range(10):
   E=(float(i)/10)*2.-1.99
   G_MATRIX=obs.conductance_matrix(MyABCwithLeads,E)

   print E,G_MATRIX[1,0]
```

The function ***conductance_matrix*** uses the function ***knitonly*** as its subroutine.
In the next section, we consider more advanced examples where the local observables for the system are calculated.

## Calculation time:

The algorithm complexity depends not only on the total system size (its number of sites) but also on the system configuration and knitting/sewing direction.
Considering the rectangular system of the size $N_{total}$, length *L* and the width *M*, as shown in fig. 9(b), we will define the interface size $M_{int}(i)$ as the maximum number of sites that at a knitting step *i* do not have all their neighbors calculated. We calculate the computational complexity for the systems in fig. 9(b) as follows:

$$CPU \propto \sum_{i=1}^{N_{total}} M_{\text{int}}(i)^2 \cong M^2(ML)$$

This calculation shows that for systems that are “longer” then “wider” the calculation time can be substantially reduced.

For the square systems with *M=L* we can also write: $CPU \propto L^4$; and for the cube with the side equal *L:* $CPU \propto L^7$.

## E. VISUALIZATION OF RESULTS

The last stage of mesoscopic system modeling with KNIT is result visualization. After the system is solved, a distribution of on-site observable parameters in the system or its transport properties can be visualized. Apart from simple textual regime, KNIT uses the capabilities of the PyX graphical package to produce plots. PyX. (http://pyx.sourceforge.net/) is a package for creating 2d and 3d plots, developed for Python. Among the main features of PyX, we can list the following:

- PostScript and PDF output for figures;
- Seamless TeX/LaTeX integration
- advanced geometric operations on paths like intersections, transformations, splitting, smoothing, etc.
- sophisticated graph generation: modular design, pluggable axes, axes partitioning based on rational number arithmetics, flexible graph styles, etc.

For more details we address the reader to the PyX documentation on the official website: http://pyx.sourceforge.net/
One can implement its own way to represent the data(for example using matplotlib package or mayavi package).

Textual: using **print a, b, c, ...**
In our example, operator **print** displays our results in a tabular form on the screen.

Graphical: using PyX and predefined functions from the library **sys**. One of the methods - ***visu2D*** **-** for system visualization was already used in this example:

```
system.visu2D(MyABCwithLeads, "<FILE NAME>")
```

Examples visualization using PyX will be explained in the next section of this document.

# 6. Advanced KNIT examples

## *6.1. Calculation of local observables for a 2D system with three leads*

In this example, we take as a basis our ABC system with the geometry as shown in fig. 6-a, 7-a :

```
import sys
sys.path.append('../')
from libpy import *

t=knit.scalarM(1.)
V=knit.scalarM(0.)
E=-0.5

""" We parametrize the system with the parameter MYSIZE = 4 """
MYSIZE=4

MyA=knit.rectangle([MYSIZE*8,4*MYSIZE],[0,3*MYSIZE],t,V)
MyB=knit.rectangle([3*MYSIZE,7*MYSIZE],[5*MYSIZE,0],t,V)
MyC=knit.rectangle([2*MYSIZE,6*MYSIZE],[0,3*MYSIZE],t,V)

MyA.coller(MyB)
MyA.coller(MyC)
```

Here we parametrise the system with MYSIZE=4 – this is equivalent to “x4 zoom” (i.e. we have 4 times more sites for the given geometry then having MSIZE=1).

We specify **three** leads and attach them to our system as follows:

```
#Making interfaces
interface_contact1=knit.rectangle([1,MYSIZE*6],[0,MYSIZE*3],t,V)
interface_contact2=knit.rectangle([MYSIZE*3,1],[5*MYSIZE,0],t,V)
interface_contact3=knit.rectangle([1,MYSIZE*2],[8*MYSIZE -1,MYSIZE*5],t,V)

#Making leads
lead1=knit.unreservoirN(interface_contact1,t)
lead2=knit.unreservoirN(interface_contact2,t)
lead3=knit.unreservoirN(interface_contact3,t)

#Attaching leads
MyABCwithLeads=knit.systemtotal(MyA,lead1)
MyABCwithLeads.addlead(lead2)
MyABCwithLeads.addlead(lead3)
```

In the resulting system we wish to make a hole. For this purposes, we specify the following method:

Initial position of a rectangular hole; its width (x-dimension) and height (y-dimension)

```
def makeHole(posX,posY, w, h):
      for x in range(w):
            for y in range(h):
                  MyABCwithLeads.move_site(
                    [MyABCwithLeads.indice([posX+x,7*MYSIZE +y-1]),
                     MyABCwithLeads.indice([posX+x-1,7*MYSIZE +y])],
                     t,V,[4*MYSIZE+x,7*MYSIZE+y],
                     MyABCwithLeads.indice([x+posX,y+posY]))
```

Here, we move sites for to the top of the system and reestablish connections between them

This method will make a rectangular hole in our system by moving sites from the central part of the system to its upper part: posX, posY – are x and y coordinates of the bottom-left corner of the hole. w, h – are width and height of the hole. We apply this method as follows:

```
makeHole(4*MYSIZE, 5*MYSIZE-1, 2*MYSIZE, MYSIZE)
```

And visualize the result:

```
import os
s=sys.argv[0]
system.visu2D(MyABCwithLeads,s+".system")
```

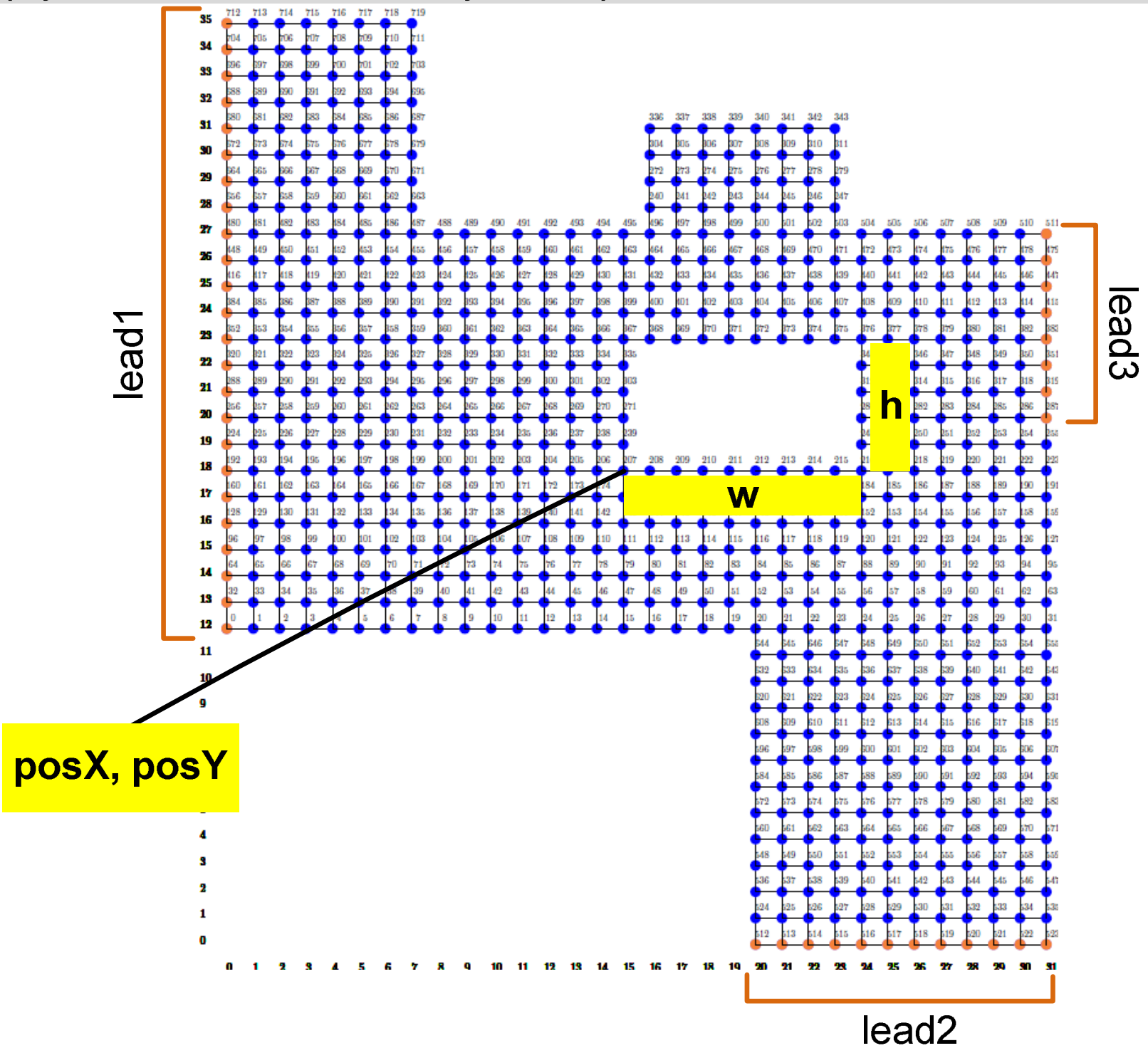

**FIG. 10:** MyABCwithLeads with 3 leads and a hole.

The method **makeHole** uses functions defined by KNIT for the system: **indice** and **move_site** . Function *indice* returns the index of the system site providing its coordinates [n1,n2,..] (in our 2D model – [x,y]). Function *move_site([v1,v2,...], hoppings, onsite_energy, [n1,n2,...], ind)* does the following: finds the system site with the index ind; moves it to the new position with coordinates [n1,n2,..] (in our 2D model – [x,y]); associates the this site with the new neighbors [v1,v2,..] and assigns new hopping and onsite energy matrices to this site. These and other KNIT functions are specified in more formal way in the next section of this document.

Now, when we have finished with the system geometry, let's have a look at the system properties.

## Introducing disorder

Until now, all the sites of the system had the same value of the on-site energy V=0. Now we introduce the disorder to our system by changing the on-site energy in a certain part of the system. To do so, we specify a method myDisorder as follows:

```
def myDisorder(W):
   import random
   random.seed(666)
   for ix in range(2*MYSIZE):
       for iy in range(2*MYSIZE):
           i=MyABCwithLeads.indice([ix+2*MYSIZE,iy+3*MYSIZE])
         MyABCwithLeads.put_Hdiag(i,knit.scalarM(W*(random.random()-0.5)))
```

This method changes the value of V for the sites in a certain square area within the system. To change the value of V for the sites, we use the method **put_Hdiag**: its first parameter is the index of a target site and the second parameter is a new value of V. This is equivalent to introducing the *impurities* into the system.

We can also visualize the resulting distribution of potential on the system as follows:

```
def potential(i):
     return MyABCwithLeads.get_Hdiag(i)(0,0).real

myDisorder(6.)

print system.colorplot2D(MyABCwithLeads,s+".pot",potential)
```

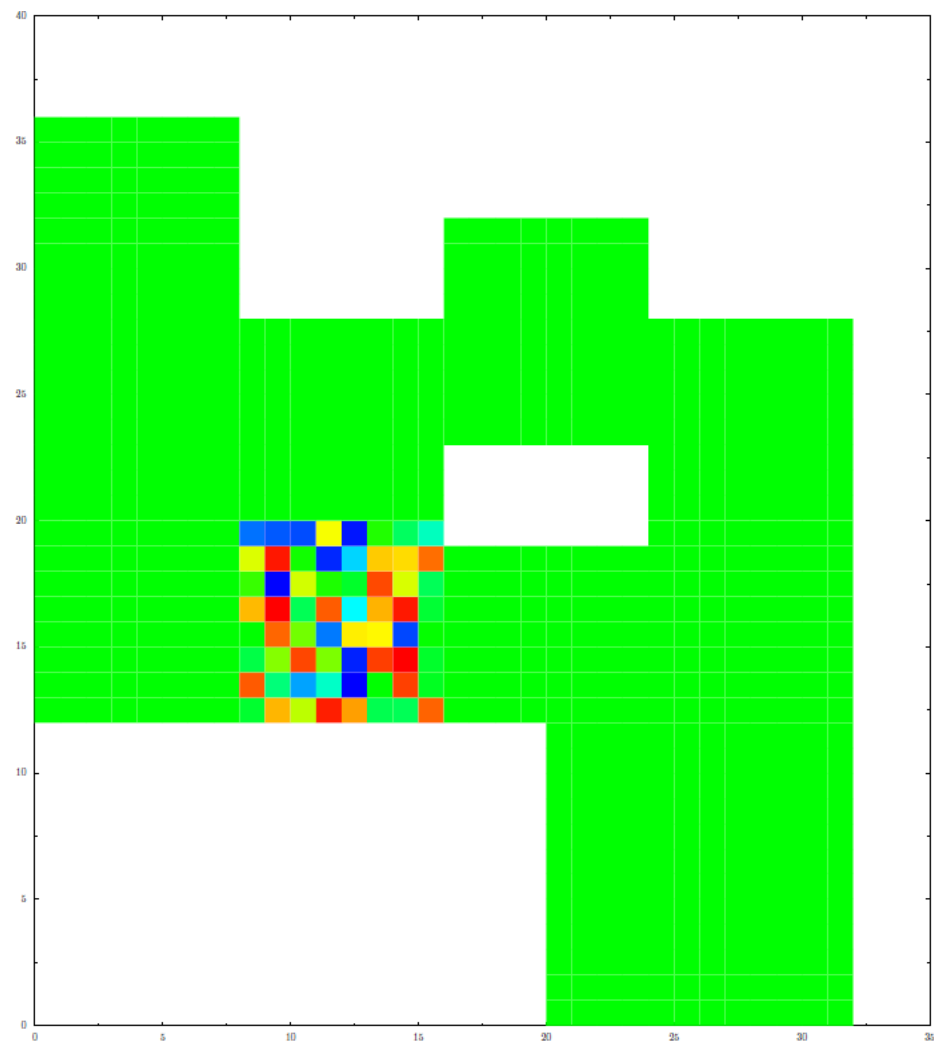

**FIG. 11:** MyABCwithLeads with introduced impurity.

The method **colorplot2D** is specified in the module system.py. This method takes as an obligatory inputs the following parameters: a pointer to the system to plot, the resulting file name, and the function that calculates the values to fill the plot area with corresponding colored squares. The color is attributed to the corresponding coordinate in the system.

**Calculation of conductance as a function of impurity strength W.**

Now we calculate the conductance properties of our system.

```
MYDATA=[]
for i in range(10):
     W=(float(i)/10)*2.
     myDisorder(W)
     G_MATRIX=obs.conductance_matrix(MyABCwithLeads,E)
     MYDATA.append([W,G_MATRIX[2,1],G_MATRIX[0,1], G_MATRIX[0,2]])
     print W,G_MATRIX[2,1],G_MATRIX[0,1],G_MATRIX[0,2]
```

Here, for each value of disorder (or impurity) strength W, we calculate the conductance as **G_MATRIX[leadIndexTo,leadIndexFrom].**

**NOTE**: the lead indexing starts from 0, i.e. the leads in fig. 10 have the following indices: lead 1 -> 0; lead 2 -> 1; lead 3 -> 2.

Therefore, G_MATRIX[2,1] calculates the conductance from the lead 2 to the lead 3 (from bottom to right), G_MATRIX[0,1] - from 2 to 1 (from bottom to left), and G_MATRIX[0,2] - from 3 to 1 (from right to left). We create the table MYDATA and fill it in based on the values of W (0..2). We also print our results on the screen.

We can also plot our results as a graph and save it into PDF file:

```
from pyx import *

g=graph.graphxy(width=10,x=graph.axis.linear(title="$W$"),
                y=graph.axis.linear(title="$G$"))
g.plot(graph.data.list(MYDATA,x=1,y=2),
           [graph.style.line([style.linewidth.Thick])])
g.plot(graph.data.list(MYDATA,x=1,y=3),
           [graph.style.line([style.linewidth.Thick,style.linestyle.dashed])])
g.plot(graph.data.list(MYDATA,x=1,y=4),
           [graph.style.line([style.linewidth.Thin,style.linestyle.dashed])])

g.writePDFfile(s+".conductance")
```

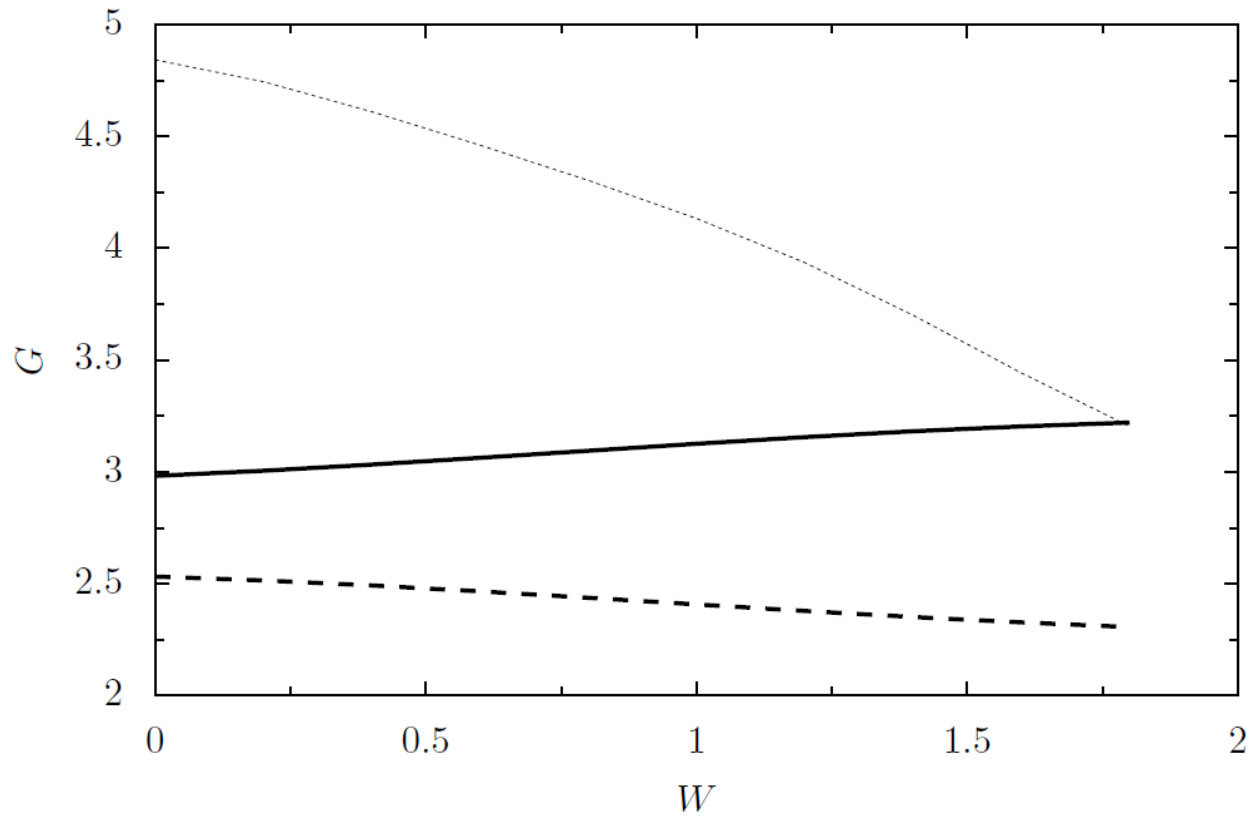

**FIG. 12: MyABCwithLeads: conductance as a function of impurity strength W**. Solid line – conductance from the lead 2 to the lead 3; Thick dashed line – conductance from the lead 2 to the lead 1; Thin dashed line – conductance from the lead 3 to the lead 1;

## Local Observables.

Finally we calculate the local observables:

```
system_solved=obs.knitANDsew(MyABCwithLeads,E)
print obs.conductance_matrix(MyABCwithLeads,E)
```

Here function **knitANDsew** executes a knitting and sewing algorithm: calculates local observables of the system; **conductance_matrix** calculates the conductance matrix for a given system.

We plot the local current coming from lead 2:

Finally we calculate all the local observables:

```
from math import *
def currentdensity(i,leadInd=1):
    tmp=0.
```

```
    for z in range(MyABCwithLeads.fzmax()):
       vi= MyABCwithLeads.voisin(i,z)
       if vi != -1 and
          MyABCwithLeads.sys2res(i)<0 and MyABCwithLeads.sys2res(vi)<0:
            Gij= system_solved.getG_lesser(i,z,leadInd)(0,0)
            Vji=MyABCwithLeads.get_Hopp(i,z)(0,0).conjugate()
            tmp=tmp+((Vji*Gij).real)**2
    return sqrt(tmp)

print system.colorplot2D(MyABCwithLeads,s+".cur",currentdensity)
```

Function **getG_lesser** returns of Lesser Green function between the site i and its z-th neighbour, given that the electrons are injected with probability 1 from the lead leadind.

Function **get_Hopp** from the module knit.py (see next section for details) returns the hopping between the site i and its z-th neighbour.

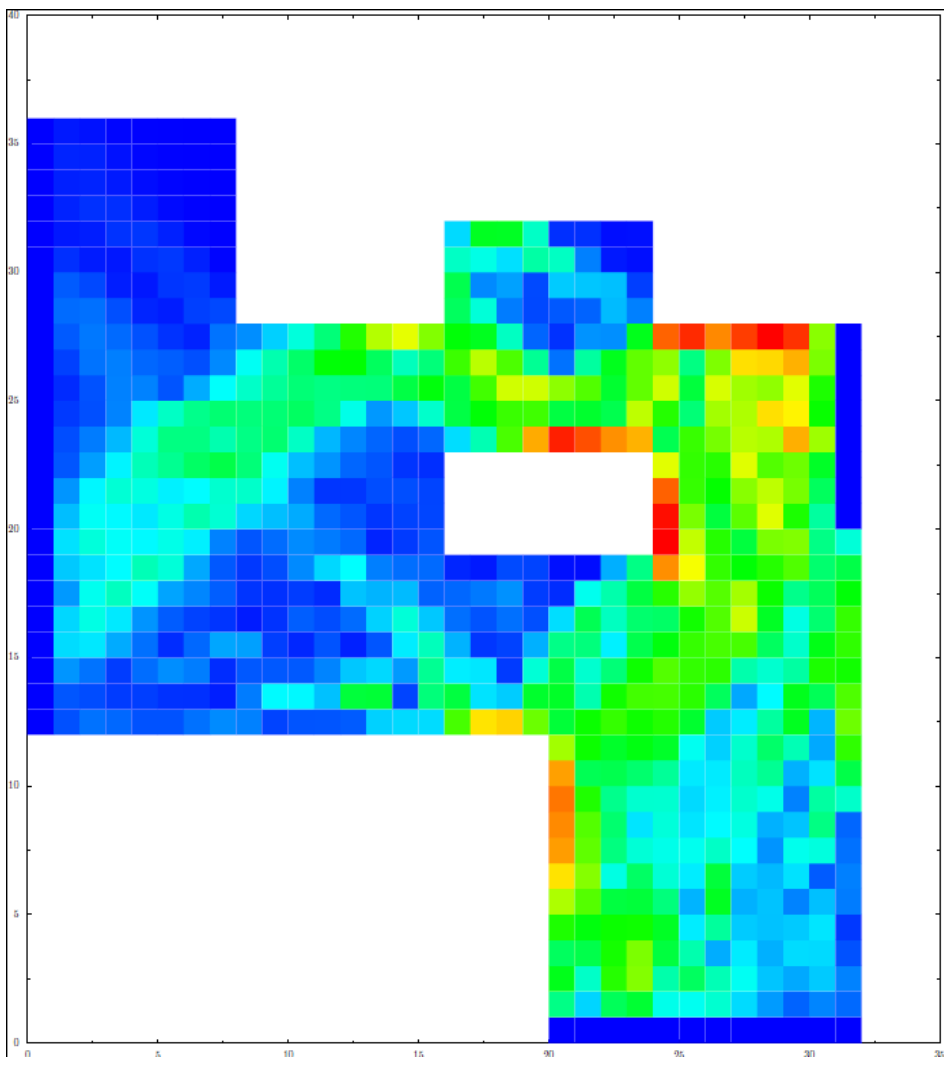

**FIG. 13: MyABCwithLeads: current from the lead 2.** Red – maximum current; Blue – no current.

## *6.2. Calculation of spin accumulation for a 3D system with spin*

In this example we consider the calculation of spin accumulation for a Py/Cu sample defined in 3D space. In this Figure 14 illustrates the configuration of the sample.

To make spin-related calculations, we will use the module spin.py:

```
import sys
sys.path.append('../')
from libpy import *
from numpy import *
from libpy.spin import *
```

We define the properties of the system with spin as follows:

```
t=spin(1.)
V=spin(0.)
E=-0.5
```

### A. System construction.

As in the previous example, we construct a system as a superposition of three rectangular blocks:

```
MYSIZE=2

A=knit.rectangle_S([MYSIZE*7,2*MYSIZE,3*MYSIZE],[0,0,0],t,V)
B=knit.rectangle_S([3*MYSIZE,10*MYSIZE, 3*MYSIZE],[2*MYSIZE,0,0],t,V)
C=knit.rectangle_S([MYSIZE*7,2*MYSIZE,3*MYSIZE],[0,8*MYSIZE,0],t,V)

A.coller(B)
A.coller(C)
```

We build the leads (also as 3D rectangles):

```
interface_contact1=knit.rectangle_S([MYSIZE*7,1,3*MYSIZE],[0,0,0],t,V)
interface_contact2=knit.rectangle_S([MYSIZE*7,1,3*MYSIZE],[0,10*MYSIZE-
1,0],t,V)

lead1=knit.unreservoirN_S(interface_contact1,t)
lead2=knit.unreservoirN_S(interface_contact2,t)

mySystem3D=knit.systemtotal_S(A,lead1)
mySystem3D.addlead(lead2)
```

**NOTE: we use functions from KNIT with a postfix "*_S*" that stands for "spin".**

To plot the 3D system function visu3D can be used. Note, however, that for systems with huge number of sites such image will be difficult to analyze.

### B. Introducing a disorder

To model properties that correspond to selected materials – **Py** and **Cu** – we add disorder to matrix elements of the corresponding sample parts.

For **Py**, the disorder is described by the following expression:

$$t_{i,j} = \begin{pmatrix} 1 & 0 \\ 0 & 1 \end{pmatrix} + W_{so} \begin{pmatrix} \xi & -\eta^* \\ \eta & \xi^* \end{pmatrix}; \quad V_i = \begin{pmatrix} W_1 \cdot \chi & 0 \\ 0 & W_2 \cdot \chi \end{pmatrix}$$

Where: $\mathrm{Re}\,\xi, \mathrm{Re}\,\eta = [-0.5..0.5]; \quad \mathrm{Im}\,\xi, \mathrm{Im}\,\eta = [-0.5..0.5]; \quad \chi = [-0.5..0.5];$

$W_{so}$=0.0; $W_1$ = 1.3, $W_2$ = 4.2

For **Cu**, the disorder is defined as follows:

$$V_i = W\begin{pmatrix} \chi & 0 \\ 0 & \chi \end{pmatrix}$$

Where: $\chi = [-0.5..0.5]$ and $W = 0.45$.

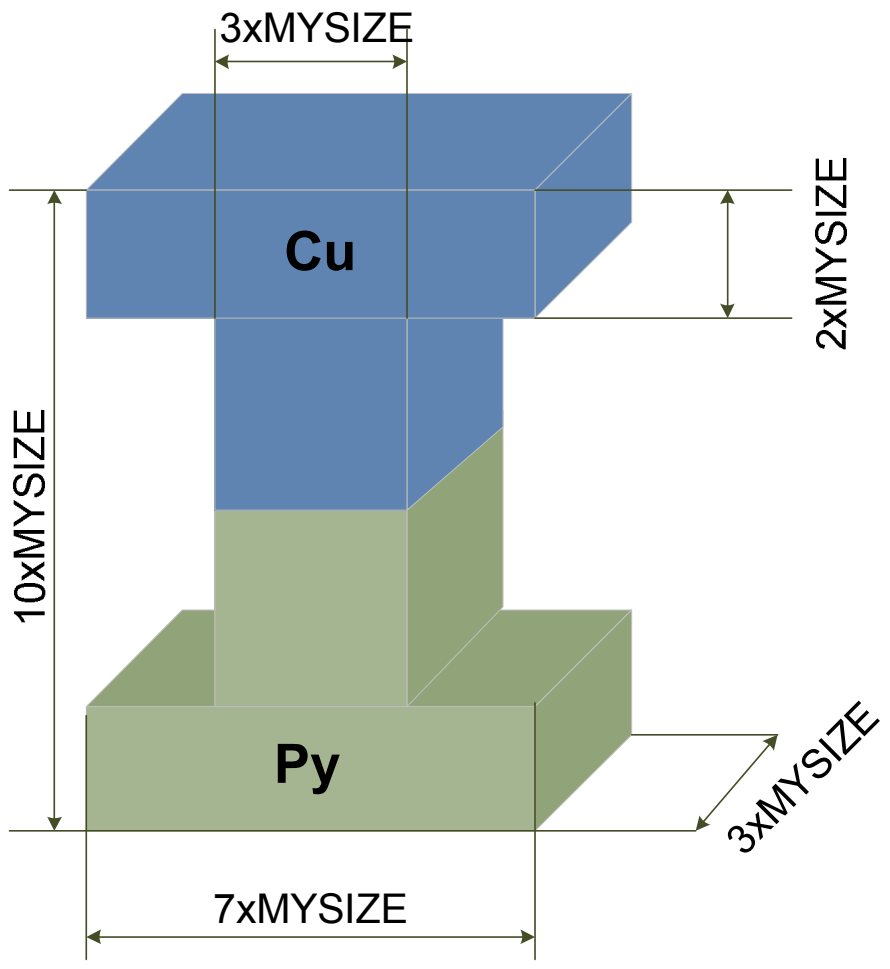

**FIG. 14:** System with 3D-geometry

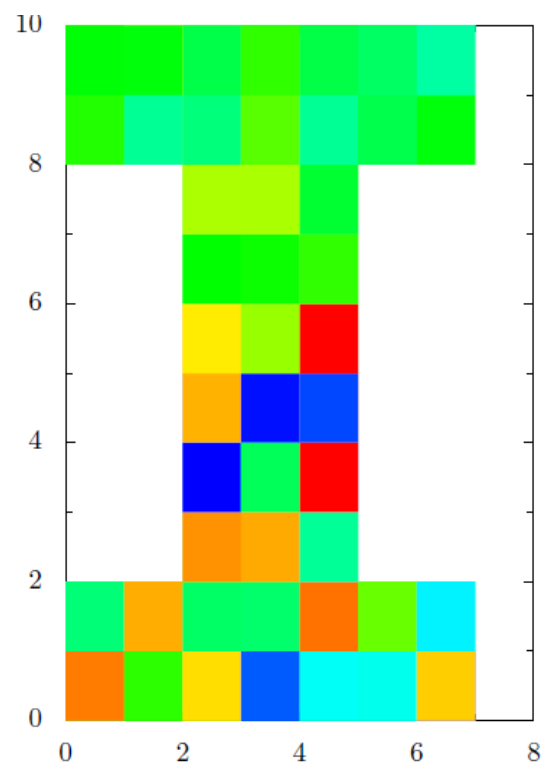

**FIG. 15:** 2D plot of the system with the parameterization MSIZE=1. Disorder corresponds to the parts made of different materials: Py (on the bottom) and Cu (on top)

First we introduce the disorder parameters in Python:

```
yMax = 10*MYSIZE
W=0.45
W1= 1.3
W2= 4.2
Wso = 0.0
```

Here yMax – is a height of our sample or the number of its cross-sections. Later we will use this parameter to normalize the conductance on each cross-section of the sample.

We specify the function that introduces disorder in our system as follows:

```
def putDisorder():
    import random
    random.seed(667)
    for i in range(mySystem3D.size()):
        y = mySystem3D.coord(i,1)  # y-coordinate of the
        if (y <= yMax/2):          # bottom part - Py
            hi = random.random()-0.5
            Vi_new = spin();
            Vi_new.set(0,0, hi*W1)
            Vi_new.set(0,1, 0)
            Vi_new.set(1,0, 0)
            Vi_new.set(1,1, hi*W2)
            mySystem3D.put_Hdiag(i,Vi_new)
            ksiRe = random.random()-0.5
            ksiIm = random.random()-0.5
            etaRe = random.random()-0.5
            etaIm = random.random()-0.5
            for z in range(mySystem3D.fzmax()):
                ti_new = spin()
                ti_new.set(0,0, Wso*(ksiRe+ksiIm*1.J))
                ti_new.set(0,1,(-1)*(Wso*etaRe-etaIm*1.J))
                ti_new.set(1,0, Wso*(ksiRe+ksiIm*1.J))
                ti_new.set(1,1,Wso*(ksiRe-ksiIm*1.J))
                ti_fin = spin(1.)+ti_new
                mySystem3D.put_Hopp(i,z,ti_fin)
        else:                       # bottom part - Cu
            hi = random.random()-0.5
            mySystem3D.put_Hdiag(i,spin(W*hi))
```

Random numbers generator with a seed parameter

For the bottom part of the sample ..

1 We create the zero-matrix for on-site energy

2 We set explicitly new values for each matrix element

3 We assign created matrix to the system site

We repeat the process 1-3 for the hopping element

We assign the same hopping matrix for all neighboring sites

We change the on-site energy for the Cu part of the sample:

**NOTE: The seed parameter of random function allows generating the same sequences of numbers; this is helpful when the same sample has to be reproduces several times. To develop several samples with the same geometry – we use different seeds.**

To illustrate the result of imposed disorder, we plot the 2D potential of the system using `colorplot2D` as in the previous example – see Fig. 15.

### C. Solving

We calculate local observables:

```
results=obs.knitANDsew_S(mySystem3D,E)
```

We define the function for accumulation count to calculate the conductance at each cross-section of the sample along the axis Y, considering the active lead -> lead 1 (Cu side):

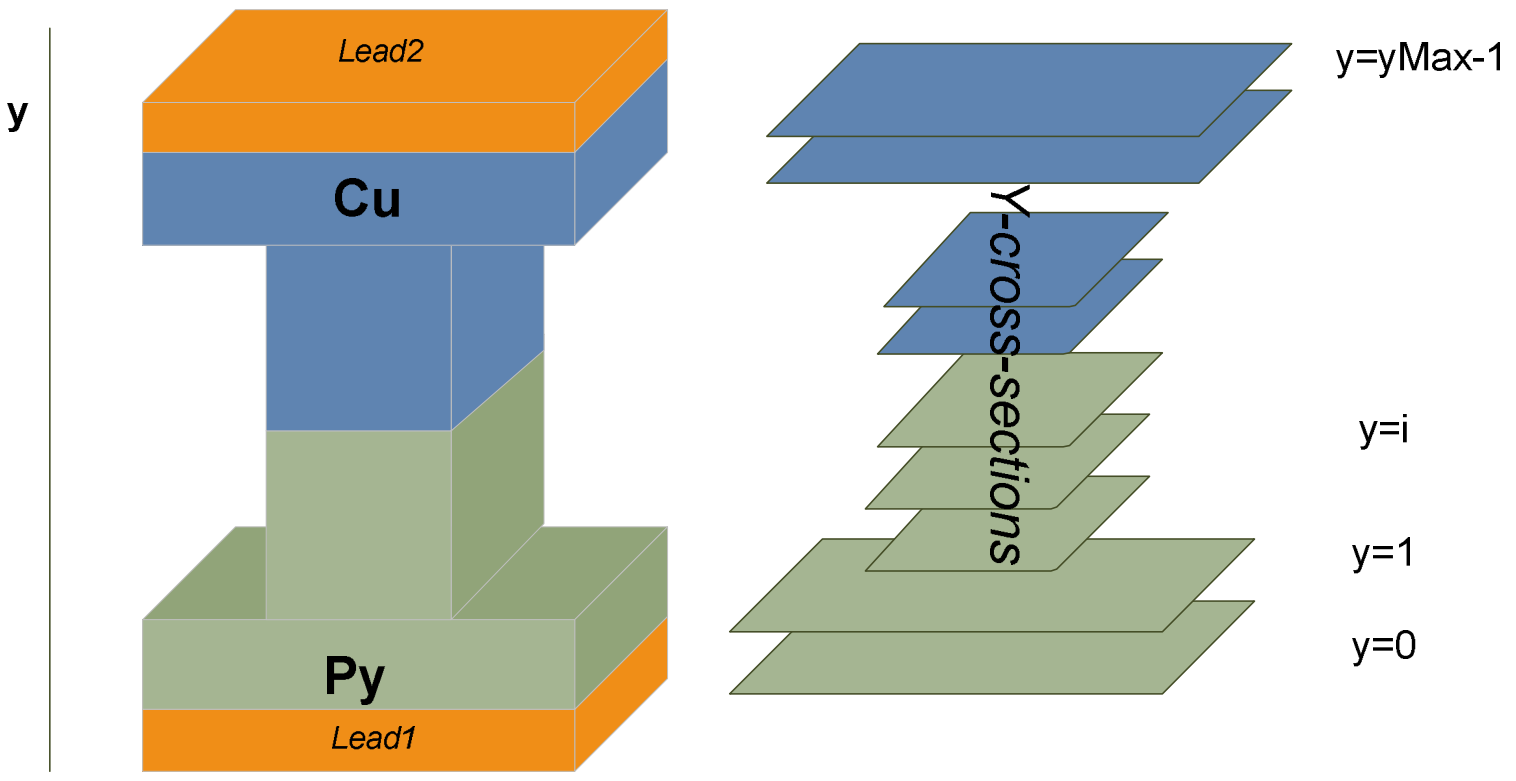

**FIG. 16:** Calculation of the conductance for y-cross-sections of the system

We define the variable to store the results for further printing

We define the accumulator as an array with elements corresponding to cross-sections

We calculate conductance for the elements belonging to the same cross-sections separately

We normalize the conductance dividing it by the number of sites in a given cross-section

```
MYDATA=[]

def getSpinAccum(li):
      zmax = mySystem3D.fzmax()
      accum=zeros((yMax,2),complex)
      for i in range(mySystem3D.fsize()):
            y = mySystem3D.coord(i,1) # y-coordinate of the site
            Gii=results.getG_lesser(i,zmax,li)(0,0)- results.getG_lesser(i,zmax,li)(1,1)
            accum[y,0]+=Gii
            accum[y,1]+=1
      res = zeros(yMax, complex)  #normalisation of result
      for j in range(yMax):
            res[j]= accum[j,0]/accum[j,1]
            MYDATA.append([j,res[j].imag])
      return res

lead=1
getSpinAccum(lead)
```

We plot the results:

```
from pyx import *

g =graph.graphxy(width=10,x=graph.axis.linear(title="$Y$"),
       y=graph.axis.linear(title="$(\mu\uparrow - \mu\downarrow)$"))
g.plot(graph.data.list(MYDATA,x=1,y=2),
       [graph.style.line([style.linewidth.Thick])])
g.writePDFfile(fName+".accum5_20samples")
```

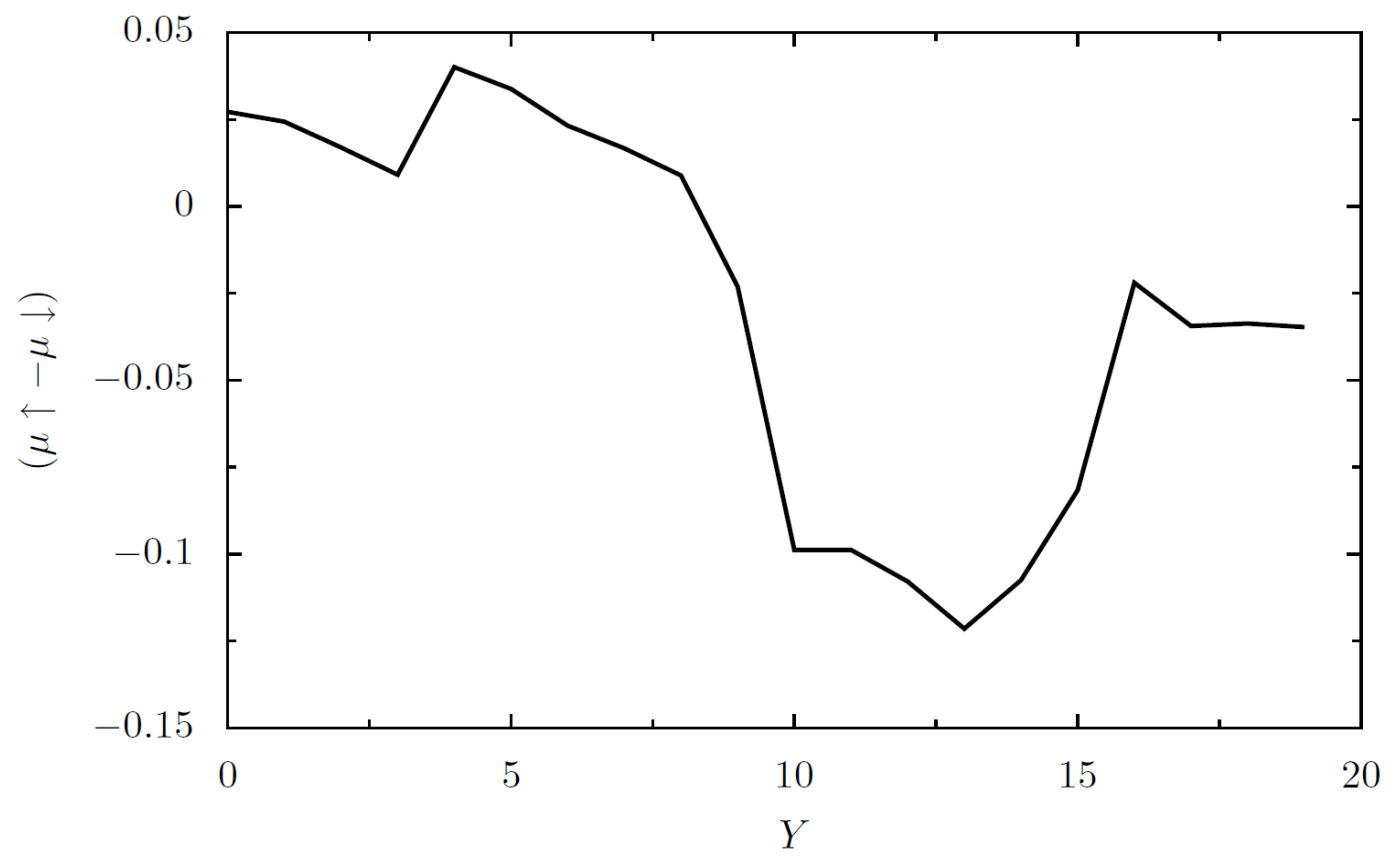

**FIG. 17:** Calculation of the conductance for Y-cross-sections of the system. This result is obtained by averaging on 20 samples. Mesoscopic fluctuations are still strong, so the curve doesn't have the smooth shape.

To obtain better approximation, one can take an average result on the set of many samples. Each sample corresponds to the different disorder; in its turn, it is simulated by changing the seed of the generator of random numbers.

## *6.2. Calculation of angular magneto-conductance in magnetic multi-layers with 3D geometry*

In this example we consider the calculation of spin accumulation for a Py/Cu sample defined in 3D space. In this Figure 18 illustrates the configuration of the sample.

We define the properties of the system with spin as follows:

```
t=spin(1.)
V=spin(0.)
E=-0.5
```

### A. System construction.

As in the previous example, we construct a system as a superposition of three rectangular blocks and attach two leads:

```
MYSIZE=2

A=knit.rectangle_S([MYSIZE*7,2*MYSIZE,3*MYSIZE],[0,0,0],t,V)
B=knit.rectangle_S([3*MYSIZE,10*MYSIZE, 3*MYSIZE],[2*MYSIZE,0,0],t,V)
C=knit.rectangle_S([MYSIZE*7,2*MYSIZE,3*MYSIZE],[0,8*MYSIZE,0],t,V)

A.coller(B)
A.coller(C)

interface_contact1=knit.rectangle_S([MYSIZE*7,1,3*MYSIZE],[0,0,0],t,V)
interface_contact2=knit.rectangle_S([MYSIZE*7,1,3*MYSIZE],[0,10*MYSIZE-
1,0],t,V)
```

```
lead1=knit.unreservoirN_S(interface_contact1,t)
lead2=knit.unreservoirN_S(interface_contact2,t)

mySystem3D=knit.systemtotal_S(A,lead1)
mySystem3D.addlead(lead2)
```

### B. Introducing a disorder

To model properties that correspond to selected materials – **Py** and **Cu** – we add disorder to matrix elements of the corresponding sample parts.

For **Py**, the disorder is described by the following expression:

$$t_{i,j} = \begin{pmatrix} 1 & 0 \\ 0 & 1 \end{pmatrix} + W_{so} \begin{pmatrix} \xi & -\eta^* \\ \eta & \xi^* \end{pmatrix}; \quad V_i = \begin{pmatrix} W_1 \cdot \chi & 0 \\ 0 & W_2 \cdot \chi \end{pmatrix}$$

Where: $\mathrm{Re}\,\xi, \mathrm{Re}\,\eta = [-0.5..0.5]$; $\mathrm{Im}\,\xi, \mathrm{Im}\,\eta = [-0.5..0.5]$; $\chi = [-0.5..0.5]$;
$W_{so}$=0.0; $W_1$ = 1.3, $W_2$ = 4.2
For **Cu**, the disorder is mad as follows:

$$V_i = W \begin{pmatrix} \chi & 0 \\ 0 & \chi \end{pmatrix}$$

Where: $\chi = [-0.5..0.5]$ and $W$ = 0.45.

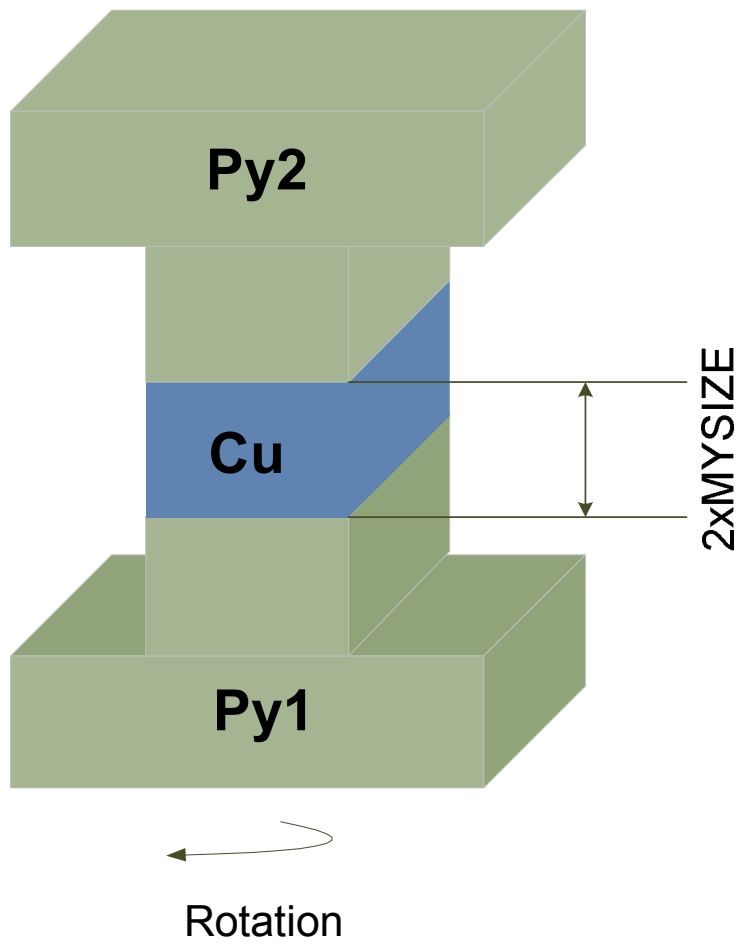

**FIG. 18:** System with 3D-geometry

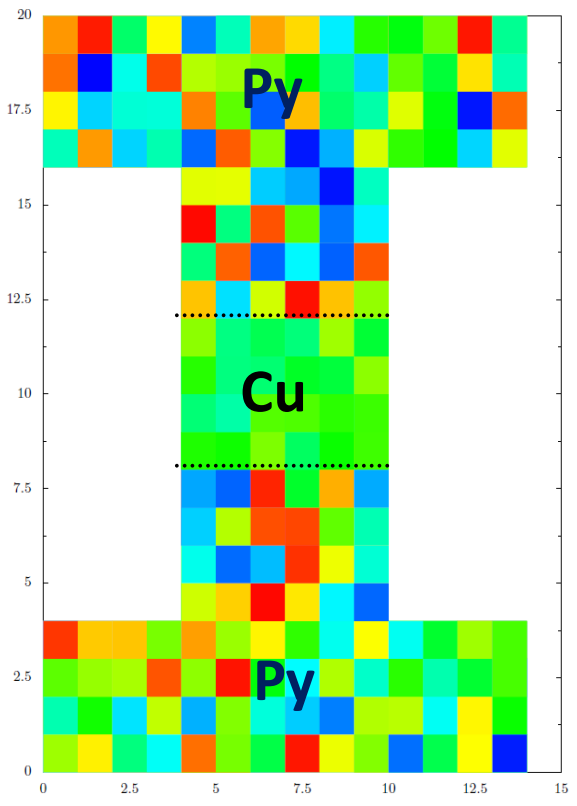

**FIG. 19:** 2D plot of the system with the parameterization MSIZE=2. Disorder corresponds to the parts made of different

materials: Py (on the bottom) and Cu (on top)

Comparing to the previous example, we have now 2 parts of Py and a middle layer of Cu. We specify the function that introduces disorder in our system as follows:

For the middle part of the sample ..

```
def putDisorder():
   import random
   random.seed(666)
   for i in range(mySystem3D.fsize()):
      y = mySystem3D.coord(i,1) # y-coordinate of the site
        if (y >= wPy1 and y < wPy1+wCu):    #  Cu
            hi = random.random()-0.5
            mySystem3D.put_Hdiag(i,spin(W*hi))
        else:                               # Py
            hi = random.random()-0.5
            Vi_new = spin();
            Vi_new.set(0,0, hi*W1)
            Vi_new.set(0,1, 0)
            Vi_new.set(1,0, 0)
            Vi_new.set(1,1, hi*W2)
            mySystem3D.put_Hdiag(i,Vi_new)
            ksiRe = random.random()-0.5
            ksiIm = random.random()-0.5
            etaRe = random.random()-0.5
            etaIm = random.random()-0.5
            for z in range(mySystem3D.fzmax()):
                  ti_new = spin()
                  ti_new.set(0,0, Wso*(ksiRe+ksiIm*1.J))
                  ti_new.set(0,1,(-1)*(Wso*etaRe-etaIm*1.J))
                  ti_new.set(1,0, Wso*(ksiRe+ksiIm*1.J))
                  ti_new.set(1,1,Wso*(ksiRe-ksiIm*1.J))
                  ti_fin = spin(1.)+ti_new
                  mySystem3D.put_Hopp(i,z,ti_fin)

putDisorder()
```

For the rest of the sample ..

### C. Solving

We calculate the conductance matrix without rotation:

```
s=obs.conductance_matrix_S(mySystem3D,E)
```

### D. Introducing magnet rotation

We define the function that rotates the bottom magnet:

```
def RotatePy1(sample, angle):
   for i in range(mySystem3D.fsize()):
      y = sample.coord(i,1) # y-coordinate of the site
        if (y >= 0 and y < wPy1):     #  Bottom part of Py
           old_Hd = sample.get_Hdiag(i)
           new_Hd = rotate(old_Hd,[0.,1.,0.],angle)
           sample.put_Hdiag(i,new_Hd)
```

Here function ***rotate***(..) is defined in the module spin.py (see Section 5); it rotates the input matrix ***old_Hd*** around one of the axes x, y, z (parameter [0,1,0] corresponds to the axis ***y*** ) on the angle defined by parameter ***angle.*** Function ***put_Hdiag*** assigns a new value of $V_{ii}$ defined by parameter ***new_Hd*** to the site ***i.***

We calculate the conductance matrix for the sample depending of the rotation angle: We turn the bottom magnet *mm* times on the *π/mm* angle to obtain the complete rotation on the angle π.  We make mm intermediate measurements and plot them.

```
mm=10
sample = mySystem3D

MYDATA=[]
phi = 0
d_phi=(float(1.)/mm)*pi #delta_phi
s=obs.conductance_matrix_S(sample,E)

MYDATA.append([phi, phi*180./pi,s[1,0]])

for i in range(mm):
    phi+=d_phi
    RotatePy1(sample, d_phi)
    s=obs.conductance_matrix_S(sample,E)
    MYDATA.append([phi, phi*180./pi,s[1,0]])
```

We plot the results:

```
from pyx import *

g =graph.graphxy(width=10,x=graph.axis.linear(title="$\phi$"),
y=graph.axis.linear(title="$g h/e^2$"))
g.plot(graph.data.list(MYDATA,x=1,y=3),[graph.style.line(
       [style.linewidth.Thick])])
g.writePDFfile(fileN+".cond")
```

For more details on plot parameters, see documentation for pyx library [4].

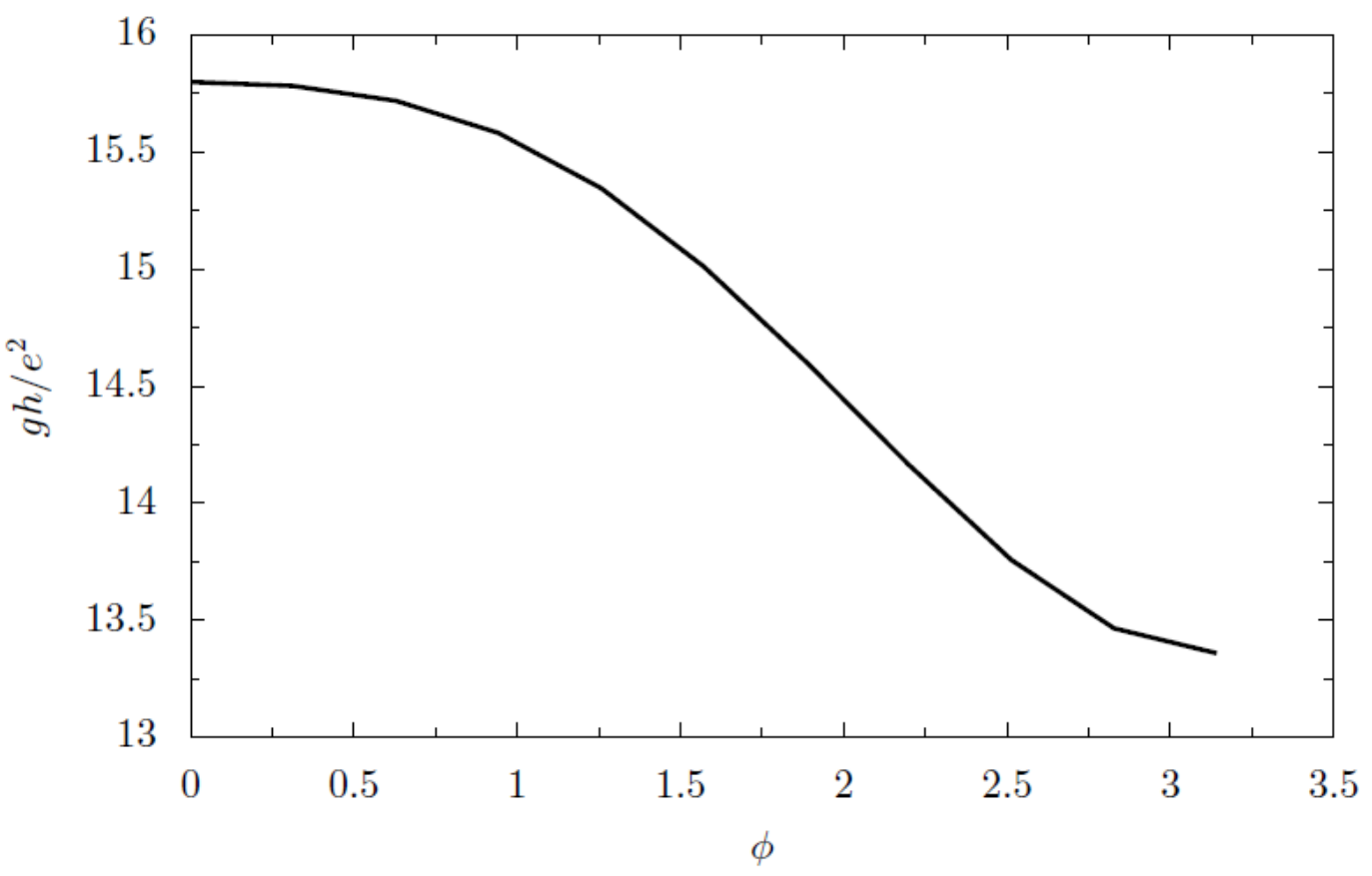

**FIG. 20:** angular magneto-resistance for a Py-Cu-Py sample

## *6.3. Calculation for a simple graphene ribbon*

In contrast to the regular (square or cubic) crystal lattices considered above, graphene requires a hexagonal crystal lattice [3]:

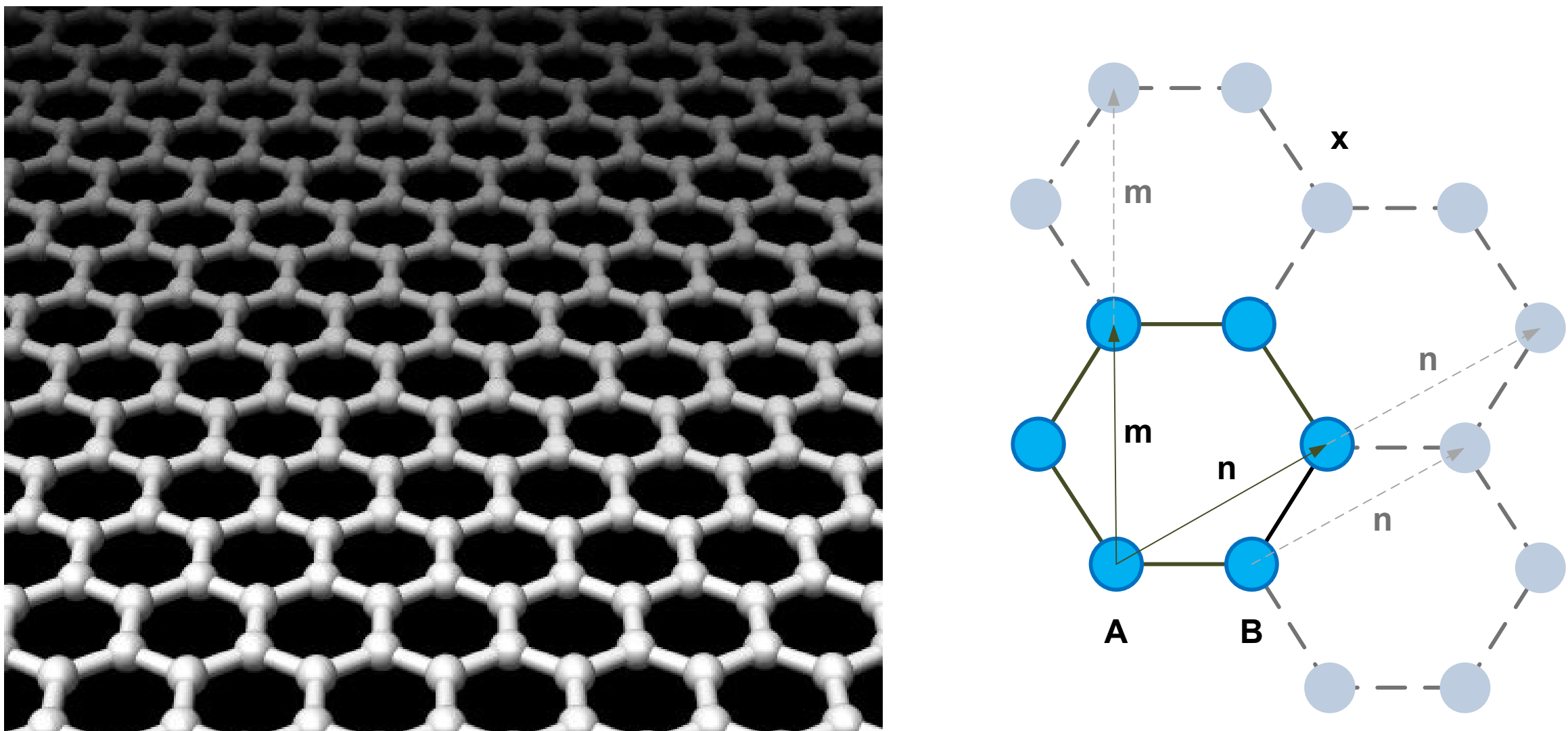

(a) (b)

**FIG. 21**: a) Hexagonal crystal lattice of graphene; b) Graphene lattice construction: A and B – two characteristic site types; n,m label the hexagons along the two directions. Each site in the ribbon can be expressed in terms of its distance from A or from B in units of m and/or n. For example, site x = A + n + m.

### A. System construction.

To implement the hexagonal lattice we introduce the following functions in Python:

```
dx,dy=2,2
coordonance=4
```

System parameterization

```
# INTRODUCING THE COORDINATE IN "GRAPHENE SPACE":
# n,m label the hexagons along the two directions
# AorB weither we are talking about a A site or a B site

# From n,m,AorB to x,y
def XY(n,m,AorB):
   if AorB=='A': AorB=0
   if AorB=='B': AorB=1
   x=3*dx*n/2+dx*AorB
   y=dy*n+2*dy*m
   return [x,y]
```

Coordinate translation: from lattice to XY

```
# From x,y to n,m, AorB
def nmAorB(x,y):
   tmp=(2*x)/dx
   n=tmp/3
   AorB=(tmp%3)/2
   m=(y-dy*n)/(2*dy)
   return n,m,AorB
```

Coordinate translation: from XY to lattice

```
def Neighbours(n,m,AorB,voisin):
   s= -2*AorB+1
   z=voisin-1
   return n-z*z*s,m+s*z*(1+z)/2,1-AorB
```

We also redefine functions to manipulate sites in graphene:

```
def add_one_site(n,m,AorB,t,Hd,monsys,x=0,y=0):
   if AorB=='A': AorB=0
   if AorB=='B': AorB=1
   X,Y=XY(n,m,AorB)
   if monsys.indice([X+x,Y+y]) != -1: return
   v=[]
   for i in range(3):
      nv,mv,AorBv=Neighbours(n,m,AorB,i)
      xv,yv=XY(nv,mv,AorBv)
      xv=xv+x
      yv=yv+y
      where=monsys.indice([xv,yv])
      if where != -1 : v.append(where)
   monsys.add_site(v,t,Hd,[X+x,Y+y])


def move_one_site(n,m,AorB,t,Hd,monsys,ind,x=0,y=0):
   if AorB=='A': AorB=0
   if AorB=='B': AorB=1
   X,Y=XY(n,m,AorB)
   v=[]
   for i in range(3):
      nv,mv,AorBv=Neighbours(n,m,AorB,i)
      xv,yv=XY(nv,mv,AorBv)
      xv=xv+x
      yv=yv+y
```

```
      where=monsys.indice([xv,yv])
      if where != -1 : v.append(where)
   monsys.move_site(v,t,Hd,[X+x,Y+y],ind)
```

We define a Zigzag constructor for a system:

To build a sample of graphene ribbon in KNIT, we are using a rectangle system constructor to allocate the memory for the needed sites, and then move the created sites to build a system with hexagonal lattice:

```
# Zigzag rectangle; lace from left to right
def ZigzagLR(nmSize,Init_xy,t,Hd):
   n_max,m_max=nmSize
   fullsize=2*n_max*m_max+2*n_max+2*m_max
   x,y=Init_xy
   monsys = knit.rectangle([fullsize,1],[x,y],t,Hd)
   vv=[]
   ind=0
   monsys.move_site(vv,t,Hd,[x,y],ind)
   # monsys=knit.rectangle([1,1],[x,y],t,Hd)
   for m in range(m_max):
      ind+=1
      move_one_site(-1,m+1,'B',t,Hd,monsys,ind,x,y)
      ind+=1
      move_one_site(0,m+1,'A',t,Hd,monsys,ind,x,y)
   for n in range(n_max):
      for m in range(m_max+1):
         ind+=1
         move_one_site(n,m,'B',t,Hd,monsys,ind,x,y)
      for m in range(m_max):
         ind+=1
         move_one_site(n+1,m,'A',t,Hd,monsys,ind,x,y)
      if n+1 != n_max:
         ind+=1
         move_one_site(n+1,m_max,'A',t,Hd,monsys,ind,x,y)
   return monsys
```

We also introduce a constructor for the lead that builds a zigzag-shaped lead :

```
# Zigzag for left or right lead
def ZigzagLeadLR(n,m_min,m_max,Init_xy,t,Hd,LR):
   #LR = O : Left lead , LR=1 : Right lead
   if LR=="RIGHT": LR=1
   if LR=="LEFT": LR=0
   x,y=Init_xy
   X,Y=XY(n,m_min,LR)
   monsys=knit.rectangle([1,1],[x+X,y+Y],t,Hd)
   for m in range(m_min+1,m_max+1):add_one_site(n,m,LR,t,Hd,monsys,x,y)
   for m in range(m_min,m_max+1):add_one_site(n,m,1-LR,t,Hd,monsys,x,y)
   a=knit.unreservoirN(monsys,knit.scalarM(0.),0.)
   L=m_max-m_min+1
   for i in range(L):
```

```
      a.put_leadHopp(i,i+L,t)
   for i in range(L-1):
      a.put_leadHopp(i+LR,i+L+1-LR,t)
   for i in range(L):a.put_leadHopp(i+L,i,knit.scalarM(1.e-5))
   return a
```

Building the system:

```
t=knit.scalarM(1.)
Hd=knit.scalarM(0.)

a=ZigzagLR([N,M],[0,0],t,Hd)
```

We add two more sites on the edges:

```
add_one_site(N,M,'A',t,Hd,a)
add_one_site(-1,0,'B',t,Hd,a)
```

We construct leads and make the total system:

```
lead0=ZigzagLeadLR(-1,0,M,[0,0],t,Hd,"LEFT")
monsys=knit.systemtotal(a,lead0)
lead1=ZigzagLeadLR(N,0,M,[0,0],t,Hd,"RIGHT")
monsys.addlead(lead1)
```

Result is plotted using standard function visu2D as explained above. The plot is illustrated in Fig. 22:

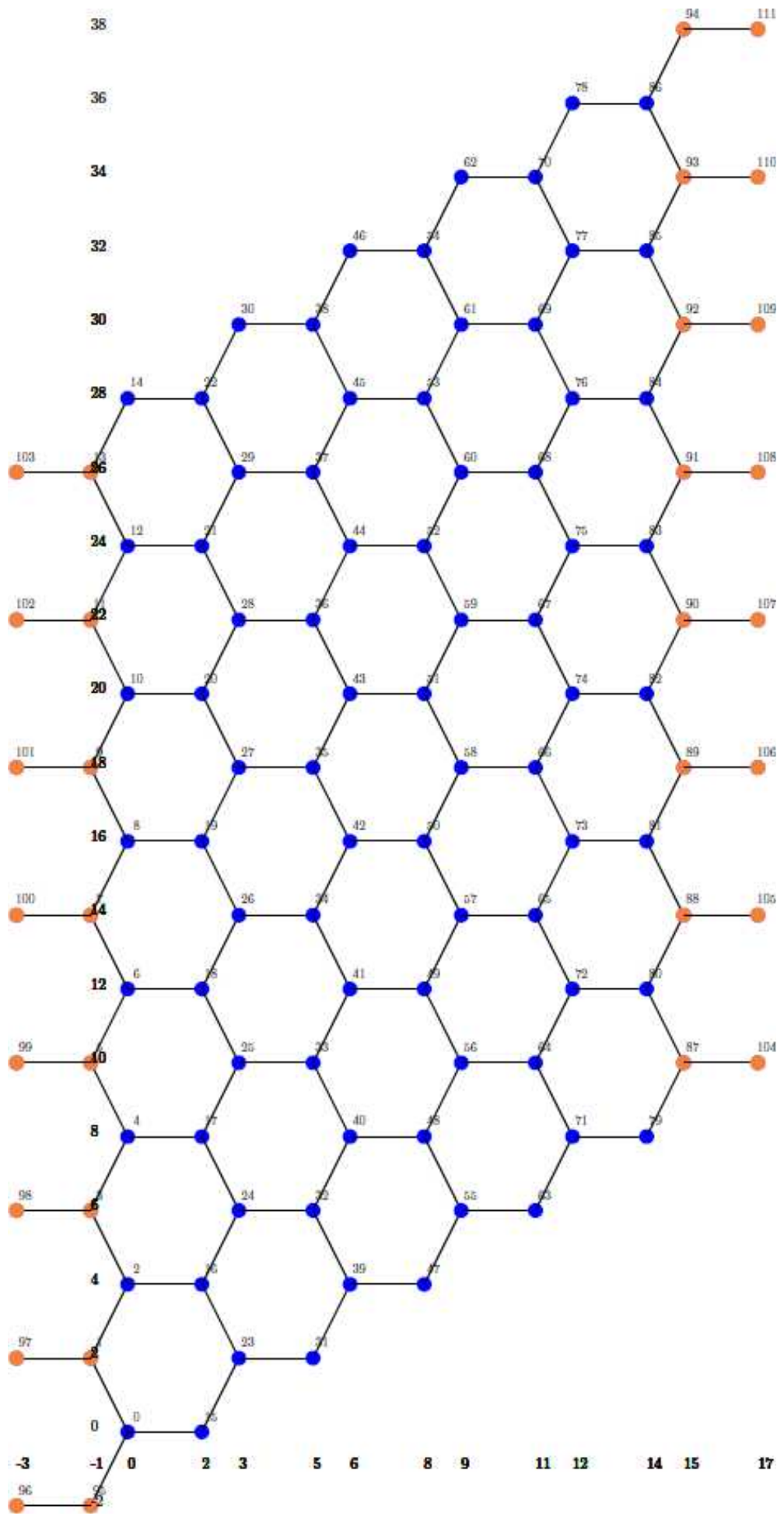

**FIG. 22**: simplest_graphene example: the system monsys plotted with visu2D.

## B. Solving

**NOTE**: With the lead solver than we are currently using, the hopping matrix ***t*** (from one layer to the next of the lead) has to be invertible. There is no fundamental reason for that and this should be fixed in later releases. As this matrix is not invertible for graphene we use a trick of adding a small matrix to t (typically 1e-5 smaller than the bandwidth). The trick works well for medium size systems but eventually become unstable.

Using KNIT, we calculate the conductance matrix for graphene ribbon and plot the result for different energies.

```
MYDATA=[]
Ndata=101
for i in range(Ndata):
     E= 3.5*((float(i)/Ndata)*2.-1.001)
     G_MATRIX=obs.conductance_matrix(monsys,E)
     MYDATA.append([E,G_MATRIX[0,1]  ])
```

```
     print E,G_MATRIX[0,1]
from pyx import *

g=graph.graphxy(width=20,x=graph.axis.linear(title="$E$"),
     y=graph.axis.linear(title="$G$"))
g.plot(graph.data.list(MYDATA,x=1,y=2))

forplot=sys.argv[0]+".res_new"+str(M)+"X"+str(N)
g.writeEPSfile(forplot)
g.writePDFfile(forplot)
```

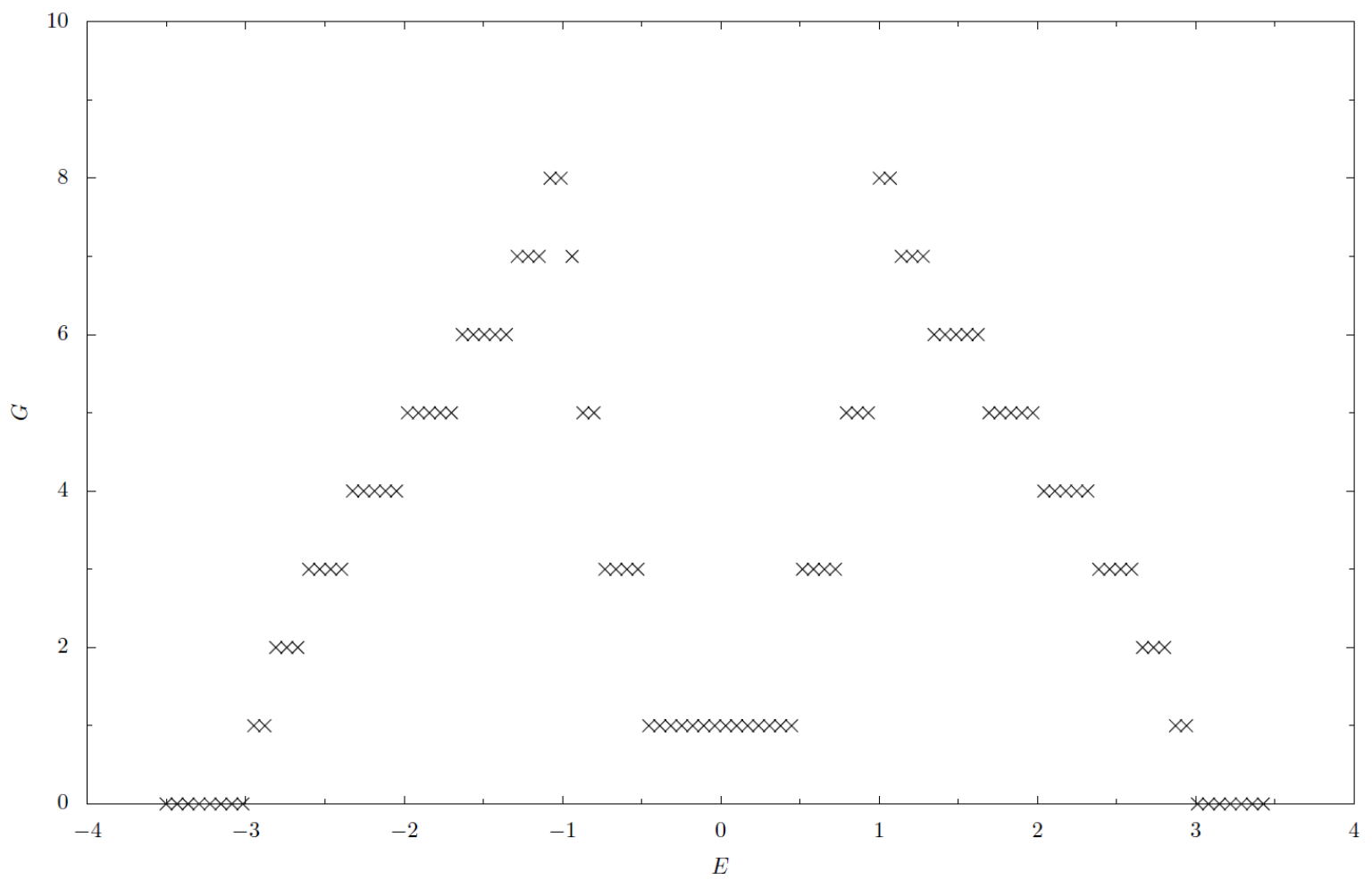

**FIG. 22**: Calculation of conductance matrix for different energies for graphene ribbon.

Our tutorial does not cover all the knit functionality. The rest of the important functions are addressed in section 5 of this document.

# 7 List of available KNIT functions and methods

KNIT user functions are defined in the svn_knit/../libpy/ directory. Three KNIT modules (knit.py, obs.py, and system.py) cover all five KNIT modeling steps defined in section 4.1:

A. System construction
B. Lead construction
C. Total system construction
D. System solving
E. Visualisation of results

In this section, we provide KNIT functions specifications for each of these steps.
Abbreviations:

**C:** - class constructor specification
**F:** - function specification
***args*** - argument list
***return:*** - return type

## 7.1 Module knit.py

This module is built by SWIG from the c++ Kernel. It provides functions to work with a system - its sites and their parameters.

## A. System construction

### Site types:

### class scalarM – no spin

scalarM is a simplest type of micromatrix . The value *v* of a scalarM micromatrix is a COMPLEX number

| | | |
|---|---|---|
| **C:** | scalarM()<br>scalarM(COMPLEX v) | creates a ZERO element v<br>creates a value v |
| *args* | *COMPLEX v* | *- value of a micromatrix; scalarM() - zero element or 0. ; scalarM(v) – value v* |
| **F:** | hconjugate() | calculates Hermitian conjugate of the value v |
| *return:* | *scalarM* | |
| **F:** | inverse(): | inverses the value v of scalarM : v = 1./v |
| *return:* | *scalarM* | |

## **class spinM** – with spin

To consider a system with spin - system sites should be defined as spinM elements. The value *v* of spinM micromatrix is a matrix 4x4 of COMPLEX numbers.

| | | |
|---|---|---|
| **C:** | vectorM()<br>vectorM(COMPLEX v) | creates a ZERO matrix v<br>creates a diagonal matrix v |
| *args* | *COMPLEX v* | *- value of a micromatrix; vectorM() - zero matrix; vectiorM(v) – diagonal matrix v* |
| **C:** | vectorM(vectorM vv) | copy constructor (v = vv) |
| *args* | *vectorM vv* | *- a micromatrix to copy from* |
| **F:** | hconjugate() | - calculates Hermitian conjugate for v |
| *return* | *spinM* | |
| **F:** | inverse(): | inverses v : v = 1./v |
| *return* | *spinM* | |
| **F:** | set(const int i,const int j,COMPLEX E) | sets the element {i,j} to E |
| *args* | *int i - row number; 0..1*<br>*int j - column number; 0..1*<br>*E - new value* | |
| *return* | *spinM* | |

## **System types:**

## **class unsysteme**

represents a system defined with scalarM micromatrix type - system without spin.

| | | |
|---|---|---|
| **C:** | unsysteme()<br>unsysteme(unsysteme s) | Creates an empty system<br>Copy constructor: creates a copy of s |
| | SYSTEM PROPERTIES: | |
| **F:** | fdim() | Returns system dimension |
| *return* | *int* | |
| **F:** | fsize() | Returns total number of sites in the system |
| *return* | *int* | |
| **F:** | fzmax() | Returns maximum possible number of neighbours for a system site |
| *return* | *int* | |
| | OPERATIONS WITH SITES: | |
| **F:** | indice(Array<int,1> coord) | Returns the site index by its coordinates given as a 1D - array *coord. (see also coord).* |
| *args* | *Array<int,1> coord* | *- similar to int [dim], where dim is a dimensionality of the system; for example, coord[0]* |
| | | *- x-th coordinate, coord[1] - y-th coordinate, coord[2] - z-th coordinate, etc. If coord.length > dim -> error!* |
| *return* | *int* | *site index.* |
| **F:** | coord(int ind, int d) | Returns the physical coordinates d of a given site ind *(see also indice).* |
| *args* | *int ind*<br><br>*int d* | *- site index. ind = 0.. size-1, where size - is a system size (total number of sites)*<br><br>*- coordinate. d= 0..dim-1, where dim is a physical dimensionality of the system. Example: d =0 -> x-th coordinate; d=1 -> y-th coordinate, etc.* |
| *return* | *int* | |
| **F:** | vcoord(int ind) | For the site with index *ind*, returns all coordinates 0..dim-1 of a given site ind in a form of 1D-array *(see also indice)* |
| *args* | *int ind* | *- site index. ind = 0.. size-1, where size - is a system size (total number of sites)* |
| *return* | *Array<int,1>* | *- an array of int with the length = dim; where dim is a dimensionality of the system. (similar to int[dim])* |
| **F:** | get_Hdiag(int ind) | For the site with index ind, returns the value of its on-site energy |
| *args* | *int ind* | *- site index. ind = 0.. size-1, where size - is a system size (total number of sites)* |
| *return* | *scalarM* | |
| **F:** | get_Hopp(int ind, int z) | For the site with index ind, returns the value of the hopping between ind and its z-the neignbour; |
| *args* | *int ind*<br>*int z* | *- site index. ind = 0.. size-1, where size - is a system size (total number of sites)*<br>*- neighbour index. z=0..fzmax-1* |
| *return* | *scalarM* | |
| **F:** | put_Hdiag(int ind, scalarM V) | For the site with index ind, sets the value of its on-site energy to Hii |
| *args* | *int ind*<br>*scalarM V* | *- site index. ind = 0.. size-1, where size - is a system size (total number of sites)*<br>*- new on-site energy* |
| *return* | **unsysteme* | *Returns a pointer to itself* |
| **F:** | put_Hopp(int ind, int z, scalarM t) | For the site with index ind, sets the value of its hopping with its z-th neighbour to t |
| *args* | *int ind*<br>*int z*<br>*scalarM t* | *- site index. ind = 0.. size-1, where size - is a system size (total number of sites)*<br>*- neighbour index. z=0..fzmax-1*<br>*- new hopping* |
| *return* | **unsysteme* | *Returns a pointer to itself* |
| **F:** | add_Hopp(int ind1, int ind2, scalarM t) | Adds a non existent hopping element t between a site with index ind1 and a site with index ind2 |
| *args* | *int ind1*<br>*int ind2*<br>*const scalarM t* | *- site index*<br>*- site index*<br>*- hopping matrix element* |
| *return* | **unsysteme* | *Returns a pointer to itself* |
| **F:** | Msite(int ind) | For the site with index ind, returns the diagonal matrix element; return zero for sites out of range |
| *args* | *int ind* | *- site index. ind = 0.. size-1, where size - is a system size (total number of sites)* |
| *return* | *scalarM* | *- diagonal matrix element - on-site energy;* |

| OPERATIONS BETWEEN SITES AND THEIR NEIGHBOURS: | | |
|---|---|---|
| **F:** | voisin(int ind,int z) | For the site with index *ind*, returns the index of its *z*-th neighbour |
| *args* | *int ind*<br><br>*int z* | *- site index. ind = 0.. size-1, where size - is a system size (total number of sites)*<br><br>*- neighbour (voisin - fr.) number. z = 0..zmax-1, where zmax - is a maximum number of neighbours for a given site.* |
| *return* | *int* | *- site index;* |
| **F:** | Mvoisin(int ind,int z) | For the site with index *ind*, returns matrix element between this site and its z-th neighbour voisin(ind,z) |
| *args* | *int ind*<br><br>*int z* | *- site index. ind = 0.. size-1, where size - is a system size (total number of sites)*<br><br>*- neighbour number. z = 0..zmax-1, where zmax - is a maximum number of neighbours for a given site.* |
| *return* | *scalarM* | *- hopping value;* |
| **F:** | vvoisin(int ind) | *Similar to voisin:* For the site with index *ind*, returns the list of indices of all neighbours in the form of 1D-array. |
| *args* | *int ind* | *- site index. ind = 0.. size-1, where size - is a system size (total number of sites)* |
| *return* | *Array<int,1>* | *- an array of int with the length = zmax ; where zmax - is a number of neighbours for a given site.* |
| **F:** | vMvoisin(int ind): | For the site with index ind, returns matrix elements between this site and all its neighbours: {voisin(ind,0),.., voisin(ind,zmax-1) |
| *args* | *int ind* | *- site index. ind = 0.. size-1, where size - is a system size (total number of sites)* |
| *return* | *Array<scalarM,1>* | *- an array of the length zmax with elements of the type scalarM; where zmax - is a number of neighbours for a given site.* |
| **F:** | z_voisin(int ind,int z) | For the initial site with index *ind* and its *z*-th neighbour (i.e. the site with index z_ind) backtracks the initial site *ind* from *z_ind* and returns the neighbour number of *ind* in the list of neighbours of the *z_ind*<br>// ind=voisin(voisin(ind,z),z_voisin(ind,z)) |
| *args* | *int ind*<br><br>*int z* | *- site index. ind = 0.. size-1, where size - is a system size (total number of sites)*<br><br>*- neighbour number. z = 0..zmax-1, where zmax - is a maximum number of neighbours for a given site.* |
| *return* | *int* | *- neighbour number; if no neighbour that has an index ind -> return z_ind* |
| SYSTEM CONSTRUCTION VIA BLOCK GLUING AND SITE MANIPULATION: | | |
| **F:** | coller(unsysteme B) | Merges itself with a system block B. Overlapped part is taken from B system: i.e. when physical coordinates of some native sites are equal to physical coordinates of some sites of B - > these native sites are replaced with the corresponding sites of B. In a resulting system sites are re-indexed to keep a continuous enumeration. |
| *args* | *unsysteme B* | *A system block to glue on top* |
| *return* | **unsysteme* | *Returns a pointer to itself* |
| *F:* | *addsite(...)* | *See add_site* |
| F: | add_site(Array<int,1> voisins,<br>Array<scalarM,1> hoppings,<br>scalarM V,<br>Array<int,1> coordinate) | adds a new site to the system with the neighbors list voisins, hopping list hoppings, onsite energy V and coordinate of this site - coordinate |
| *args* | *Array<int,1> voisins*<br>*Array<Tmicromatrix,1> hoppings,*<br>*Tmicromatrix V,*<br>*Array<int,1> coordinate,* | *- list of neighbour indices*<br>*- list of hoppings*<br>*- on-site energy*<br>*- physical coordinate* |
| *return* | **unsysteme* | *Returns a pointer to itself* |
| **F:** | move_site(Array<int,1> voisins,<br>Array<scalarM,1> hoppings,<br>scalarM V,<br>Array<int,1> coordinate,<br>int iii) | moves a site iii to a new coordinate coord by assigning a new neighbour list voisins, hoppings from this site, and onsite energy V |
| *args* | *Array<int,1> voisins*<br>*Array<scalarM,1> hoppings,*<br>*scalarM V,*<br>*Array<int,1> coordinate,*<br>*int iii* | *- list of neighbours*<br>*- list of hoppings*<br>*- on-site energy*<br>*- physical coordinate*<br>*- site index* |
| *return* | **unsysteme* | *Returns a pointer to itself* |
| *F:* | *movesite(...)* | *See move_site* |
| **F:** | shift_system(Array<int,1> delta) | Shifts the hole system on the vector delta in coordinate space. All the site properties rest the same. |
| *args* | *Array<int,1> delta* | *- shift vector* |
| *return* | **unsysteme* | *Returns a pointer to the shifted system* |

## class rectangle - *extends unsysteme*

This class defines a rectangular system without spin. All functions of unsysteme are also defined for rectangle. Supplementary functions are defined below:

| | | |
|---|---|---|
| **C:** | rectangle(Array<int,1> geometry, Array<int,1><br>bottom_left_corner,<br>scalarM t, scalarM V) | Constructor from size, hopping and on-site energy |
| *args* | *Array<int,1> geometry*<br><br>*Array<int,1> bottom_left_corner*<br><br>*scalarM t*<br>*scalarM V* | *- physical size of the block given as an Array of integers: [width, height (,depth - for 3D-geometry)];*<br>*- absolute coordinates of the bottom left corner of the block given as an Array of integers: [x0, y0 (, z0 - for 3D case)]*<br>*- hopping matrix*<br>*- on-site energy matrix* |

## class unsysteme_S, class rectangle_S - *extends unsysteme_S*

Represent a system with a spin and a rectangular system with a spin. (i.e. micromatrix type is spinM). All functions, defined for the system without spin (see class unsysteme above)are also valid for the system with spin assuming that all parameters of the type scalarM are replaced with the corresponding parameters of the type spinM.

# B. Leads construction

## class unreservoir - *extends unsysteme*

| | | |
|---|---|---|
| **F:** | green(COMPLEX E) | Calculates a Green function of a lead interface for the energy E |
| **F:** | selfE(COMPLEX E) | Calculates a self energy of a lead interface for the energy E (supposing the lead is connected with a system via inter-slice hopping) |
| **F:** | G_SE (COMPLEX E) | Calculates a pair <Green Function , Self Energy> of a lead |

## class unreservoirN - *extends unreservoir*

Numerical reservoir (lead) for a system without spin - scalarM type of the micromatrices; extends the class unreservoir.

| | | | |
|---|---|---|---|
| **C:** | unreservoirN(unreservoirN  R) | | - a copy constructor |
| **args* | *unreservoirN  R* | *- a numerical reservoir to copy* | |
| **C:** | unreservoirN(unsysteme i, scalarM t ) | | - constructs a semi-infinite lead from the interface part i and a hopping matrix t between a system and a lead |
| **args* | *unsysteme i* | *- an interface between the system and a lead, typically defined as a system with only one layer of sites in one of its dimensions* | |
| | *scalarM t* | *- a hopping matrix between a system and a lead* | |
| **F:** | green(COMPLEX E) | Calculates a Green function of a lead interface for the energy E | |
| **args* | *COMPLEX E - value of energy* | | |
| *return* | *GF_type -* | *- a Green function, represented by an Array<COMPLEX,4>(*)* | |
| **F:** | selfE(COMPLEX E) | Calculates a self energy of a lead interface for the energy E (supposing the lead is connected with a system via inter-slice hopping) | |
| **args* | *COMPLEX E - value of energy* | | |
| *return* | *GF_type* | *- a Green function, represented by an Array<COMPLEX,4>(*)* | |

| | | |
|---|---|---|
| **F:** | get_leadHopp (int from, int to) | Returns an inter-slice hopping in the lead |
| **args* | *int from* - *index of a slice in the lead*<br>*int to* - *index of a slice in the lead* | |
| *return* | *scalarM* - *hopping* | |
| **F:** | put_leadHopp(int from, int to, scalarM t) | Sets an inter-slice hopping *t* in the lead between the slices *from* and *to* |
| **args* | *int from* - *index of a slice in the lead*<br>*int to* - *index of a slice in the lead*<br>*scalarM t* - *hopping* | |
| *return* | *unreservoirN* - *a pointer to itself* | |

(*)*GF_type* defines a Green Function type as a data structure with 4 components: GF(i,alpha,j,beta); Each component is a complex number. *GF_type – a scalar value for the system without spin*

## C. Total system

### class systemtotal - *extends unsysteme*

Describes an infinite system with leads.

| | | |
|---|---|---|
| **C:** | systemtotal(unsysteme s, unreservoir lead) | Constructs an infinite system with a single lead based on *s* and the lead *lead* |
| **args* | *unsysteme s* - *a finite system (without spin)*<br>*unreservoir lead* - *reservoir* | |
| **C:** | systemtotal(systemtotal ss) | Copy constructor |
| **args* | *systemtotal ss* - *an infinite system (without spin) to copy from* | |
| | INFINITE SYSTEM CONSTRUCTION BY ADDING LEADS: | |
| **F:** | addlead(unreservoir lead) | Connects a lead to the total infinite system |
| **args* | *unreservoir lead* - *lead* | |
| *return* | *systemtotal* - *a pointer to itself* | |
| | INFINITE SYSTEM PROPERTIES: | |
| **F:** | fNres() | Returns number of leads |
| *return* | *int* - *number of leads of the system* | |
| **F:** | p_lead(int leadN) | For a given lead number *leadN*, returns a pointer to this lead |
| **args* | *int leadN* - *lead number* | |
| *return* | *unreservoir** - *a pointer to a lead* | |
| **F:** | lead_size (int leadN) | For a given lead number leadN, returns a size of this lead - number of its sites |
| **args* | *int leadN* - *lead number* | |
| *return* | *Int* - *lead size* | |
| **F:** | max_lead_size() | Returns a maximum lead size among all connected to the system |
| *return* | Int | *- max lead size* |
| | OPERATIONS BETWEEN LEAD SITES AND THEIR NEIGHBOURS: | |
| **F:** | lead_voisin(int leadN, int index_lead, int z) | Similarly to the function *voisin* defined in the class *unsysteme,* for the lead number *leadN* and its site with index *index_lead*, this function returns the index of its *z*-th neighbour if it also belongs to this lead. Otherwise returns "not_lead". |
| **args* | *int leadN*<br>*int index_lead*<br>*int z* | *- lead number (0..Nrez-1)*<br>*- index of a lead site (0.. lead_size(lead)-1)*<br>*- neighbour number. z = 0..zmax-1, where zmax - is a maximum number of neighbours for a given site.* |
| *return* | *int* | *- index of a lead site* |
| **F:** | res2sys(int leadN,int index_lead) | Interface sites are sites that belong both to a lead and to the system; these sites have double indexation. For the lead number *leadN* and a lead site with index *index_lead*, this function returns the corresponding index of this site in the system . |
| **args* | *int leadN*<br>*int index_lead* | *- lead number (0..Nrez-1)*<br>*- index of a lead site (0.. lead_size(lead)-1)* |
| *return* | *int* | *- corresponding index of a system site* |
| **F:** | sys2res(int ind) | Interface sites are sites that belong both to a lead and to the system; these sites have double indexation. For a system (interface)site with index *ind* this function returns the lead number to which this site is connected; returns "*not_lead*" if given system site does not belong to any lead |
| **args* | *int ind* | *- index of a system site* |
| *return* | *int* | *- lead number (0..Nrez-1) or "not_lead"* |
| **F:** | sys2res_ind(int ind) | Interface sites are sites that belong both to a lead and to the system; these sites have double indexation. For a system site with index *ind*, this function returns the corresponding index of this site in the lead. *(see also res2sys)* |
| **args* | *int ind* | *- index of a system site* |
| *return* | *int* | *- corresponding index of a lead site* |

## class unreservoir_S - *extends unsysteme_S,* class unreservoirN_S - *extends unreservoir_S*

These classes are modifications of unreservoir and unreservoirN for the system with spin. All functions, defined for the unreservoir and unreservoirN above are also valid for

unreservoirN_S assuming that all parameters of the type scalarM are replaced with the corresponding parameters of the type spinM.

### class systemtotal_S *extends unsysteme_S*

This class describes an infinite system with leads and spin. As above, all functions, defined for the systemtotal are valid for systemtotal_S, considering scalarM -> spinM; unreservoirN -> unreservoirN_S; unsysteme -> unsysteme_S, etc.

## D. System solver

### class tricot

This class contains functions for realisation of the knitting algorithm for global transport properties of the system.

| | | |
|---|---|---|
| **F:** | Energy() | Returns the energy value |
| *return* | *COMPLEX* | *- energy value* |
| **F:** | set_saving_level(int level) | |
| **F:** | get_saving_level() | |
| **F:** | forward(int move_type) | Adds sites to the knitting thread |
| **args* | *int move_type* | *- defines the way of adding sites. For the thread between sites i and j (where i<j), we can make four moves:*<br>*0 = " le_debut" - add one site to the beginning of the thread (i-1);*<br>*1 = " la_fin" - add one site to the end of the thread (j+1);*<br>*2 = "ToEnd" - move to the End of the thread by adding all sites j+1, .., N-1)*<br>*3 = "ToZero" - move to the Beginning of the thread by adding all sites i-1,..,0* |
| **F:** | backward(int move_type) | Removes sites from the knitting thread (see forward) |
| **args* | *int move_type* | *- defines the way of unknitting sites. For the thread between sites i and j (where i<j), we can make four move types:*<br>*0 = " le_debut" - remove one site from the beginning of the thread (i);*<br>*1 = " la_fin" - remove one site from the end of the thread (j);*<br>*2 = "FromEnd" - if j=N-1 - removes all sites of the thread starting from j*<br>*3 = "FromZero" - if i = 0 removes all sites of the thread starting from i* |
| **F:** | G_LL(int li,int lj) | Calculates a contiguously ordered full Green function between lead li lead lj (amplitudes of transition probability) |
| **args* | *int li, lj* | *- indices of leads* |
| *return* | *GF_type* | *- Green function* |
| **F:** | G(int sys_i,int sys_j) | Calculates a Green Function element for system sites i, j |
| **args* | *int i, j* | *- indices of system sites* |
| *return* | *Array<COMPLEX,2>* | *- Green function element* |
| **F:** | G() | Calculates full GF for energy E |
| *return* | *GF_type* | *- Green function* |
| **F:** | getG_LL(int i1,int li1,int i2,int li2) | Returns retarded Green function between the sites i1 and i2 belonging to the leads li1 and li2 respectively. $G_{ij}$ |
| **args* | *int i1, i2*<br>*int li1, li2* | *The indices of lead sites: i1=0.. lead_size(li1)-1, i2=0.. lead_size(li2)-1;*<br>*Leads 0..fNres* |
| *return* | *GF_type* | *- Retarder Green function* |
| **F:** | getSE(int i,int j,int li) | For the lead number li, and sites i, j of this lead, returns self energy $\Sigma_l$ |
| **args* | *int i, j*<br>*int li* | *- indices on sites on the lead li: i,j=0.. lead_size(li)-1;*<br>*- index of a lead:  li = 0..fNrez - 1* |
| *return* | *scalar* | *- value of self energy* |
| **F:** | p_system() | Returns the pointer to the total system |
| *return* | *systemtotal** | *- system with leads* |
| **F:** | SE(int li) | For the lead number li, returns value of a contiguously ordered self energy |
| **args* | *int ln* | *- index of a lead* |
| *return* | *GF_type* | *- Value of self energy on the lead* |

(*)*GF_type* defines a Green Function type as a data structure with 4 components: GF(i,alpha,j,beta); Each component is a complex number.

## class couture

This class contains functions for realisation of the sewing algorithm for calculation of local observables of the system.

| | | |
|---|---|---|
| **F:** | sew2end() | Implement the sewing algorithm for the all system; calculates local observables of the system. We recommend the use of Python interface for this function – functions knitANDsew and knitANDsew_S from the module ***obs.py*** |
| **F:** | getG_lesser(int ind,int z,int li) | Returns the value of Lesser Green function between the site ind and its z-th neighbour, given that the electrons are injected with probability 1 from the lead li: $G_l^<$ (*) |
| **args* | *int ind*<br>*int z*<br>*int li* | *The site index, which corresponds to i from Eq.(*);*<br>*The z-th neighbour that is related to i from Eq.(*) as: j=voisin(ind, z)*<br>*Index of the lead; corresponds to l from Eq.(*);* |
| *return* | *GF_type* | *Green function* |
| **F:** | getG_LL(int i1,int lead1,int i2,int lead2) | Returns full Green Function between lead1 and lead2 with lead site indices i1 and i2 |
| **args* | *int i1, i2*<br>*int lead1, lead2* | *The indices of lead sites: i1=0.. lead_size(lead1)-1, i2=0.. lead_size(lead2)-1;*<br>*Leads 0..fNres* |
| *return* | *GF_type* | *Green function* |
| **F:** | getG_retarded(int ind,int z) | Returns Green Function (not Keldysh) for the site ind and its neighbour z (i,zmax)=on_site $G_{ij}$ |
| **args* | *int ind*<br>*int z* | *System site index; ind =0..fsize-1*<br>*Neighbour index; z = 0..fzmax; for z = fzmax function returns the on-site GF* |
| *return* | *GF_type* | *-Green function* |
| **F:** | getGnotK() | Returns full system GF (not Keldysh): GnotK(si,ALL,z,ALL) - (si,z) full GF element, (i,zmax)=on_site |
| *return* | *scalarM* | *- value of self energy* |
| **F:** | getSE(int i,int j,int li) | For the lead number li, and sites i, j of this lead, returns self energy $\Sigma_l$ |
| **args* | *int i,j*<br>*int li* | *The indices of lead sites: i,j=0.. lead_size(li)-1;*<br>*Leads 0..fNres* |
| *return* | *scalarM* | *- value of self energy* |

## class tricot_S, class couture_S

These classes are modifications of tricot and couture for the system with spin.

## 7.2 Module spin.py

These module contains simplified Python functions for working with system micromatrices – scalar and spin (see example 4.3).

| | | |
|---|---|---|
| **C:** | spin()<br>spin(COMPLEX E) | Constructs a zero element spinM (equivalent to vectorM(0))<br>Constructs a diagonal element spinM ( equivalent to vectorM(E)) |
| *args* | *COMPLEX E* | *- value of a micromatrix; vectorM() - zero matrix; vectiorM(v) – diagonal matrix v* |
| **F:** | set(int row, int col, complex value) | Sets the element [row,col] of the matrix element to *value* |
| **args* | *int row, col*<br>*complex value* | *- indices for spinM; row, col = 0,1*<br>*- value* |
| **F:** | rotate(spin element, Array<int,1> axe, float angle) | Rotates the spin element around the axe *axe* on the angle *angle (see example 4.3)* |
| **args* | *spin element,*<br>*Array<int,1> axe,*<br>*float angle* | *- micromatrix element to rotate*<br>*- axe in the form: [1,0,0] -> x; [0,1,0] ->y; [0,0,1]->z*<br>*- angle in radians* |
| *return* | *spin* | *- rotated micromatrix element* |

## 7.3 Module obs.py

This module contains facilities to call the knitting kernel and compute observables

## D. System solver

| | | |
|---|---|---|
| **F:** | conductance_matrix(systemetotal system, COMPLEX Energy) | For a given system, calculates the conductance matrix for the scalar (1-dim) micromatrices; |
| **args* | *systemetotal system - an infinite system with leads*<br>*COMPLEX Energy - energy* | |
| *return* | *Float[] -matrix of the size fNres x fNres. fNres is a number of leads in the system* | |
| **F:** | knitonly(systemetotal system, COMPLEX Energy, int savinglevel=-1) | Executes a knitting algorithm and calculates transport between leads (but not local observables) |
| **args* | *savinglevel - level of saving*<br>*-1 - exact version, saves full GF at each step – guarantees precision but memory consuming; (default value)*<br>*0- no saving; useful only for transmission calculations or very (knit only)*<br>*n - Saves full GF after each steps. n should be order of M, where M is a maximum sewing interface size.* | |
| *return* | *tricot - pointer to the data structure with solved system* | |
| **F:** | knitANDsew(systemetotal system, COMPLEX Energy, int savinglevel=-1) | Executes a knitting and sewing algorithm: calculates local observables of the system |
| **args* | *savinglevel -* | |
| *return* | *tricot - pointer to the data structure with solved system* | |
| **F:** | conductance_matrix_S(systemetotal_S system, COMPLEX Energy) | For a given system calculates the conductance matrix for the 2x2 (2-dim) micromatrices, i.e. systems with spin; |
| **args* | *Systemetotal_S system - an infinite system with leads(and with spin)*<br>*COMPLEX Energy - energy* | |
| *return* | *float[] - matrix of the size fNres x fNres. fNres is a number of leads in the system.* | |
| **F:** | knitonly_S(systemetotal_S system, COMPLEX Energy, int savinglevel=-1) | Implements knitonly (see above) for systems with spin |
| **args* | *savinglevel - level of saving*<br>*-1 - exact version, saves full GF at each step – guarantees precision but memory consuming; (default value)*<br>*0- no saving; useful only for transmission calculations or very (knit only)*<br>*n - Saves full GF after each steps. n should be order of M, where M is a maximum sewing interface size.* | |
| *return* | *pointer to the data structure with solved system* | |
| **F:** | knitANDsew_S(systemtotal_S system, | Implements knitANDsew (see above) for |
| | COMPLEX Energy, int savinglevel=-1) | systems with spin |
| **args* | *savinglevel -* | *level of saving*<br>*-1 - exact version, saves full GF at each step – guarantees precision but memory consuming; (default value)*<br>*0- no saving; useful only for transmission calculations or very (knit only)*<br>*n - Saves full GF after each steps. n should be order of M, where M is a maximum sewing interface size.* |
| *return* | | *pointer to the data structure with solved system* |
| **F:** | Green(unsysteme /_S/ lead1, unsysteme/_S/ lead2, int sigma, int eta, tricot/_S/knitted_system, systemtotal /_S/s_total) | Calculates retarded Green function $G^{\sigma\eta}_{a_k,b_l}$, where $a_k$ and $b_l$ are sites that belong to the interfaces between a system and leads k and l respectively |
| **args* | *unsysteme/unsysteme_S lead1*<br>*unsysteme/unsysteme_S lead2*<br>*int sigma*<br>*int eta*<br>*tricot/tricot_S knitted_system*<br>*systemtotal/systemtotal_S s_total-* | *– lead k*<br>*– lead l*<br>*– spin degree of freedom = 1..m*<br>*- spin degree of freedom = 1..m*<br>*- result of knitting; pointer on "kniter"*<br>*- the system* |
| *return* | *COMPLEX[]* | *A matrix of the size m x n, where m = s_total. lead_size(lead1), n = s_total. lead_size(lead2)* |
| **F:** | SelfEnergy(unsysteme /_S/ lead, int sigma, int eta, tricot/_S/knitted_system, systemtotal /_S/s_total) | Calculates self-energy $\Sigma^{\sigma\eta}_{a_l b_l}$ where $a_l$ and $b_l$ are sites of the interface between a system and a lead l . |
| **args* | *unsysteme/unsysteme_S lead*<br>*int sigma*<br>*int eta*<br>*tricot/tricot_S knitted_system*<br>*systemtotal/systemtotal_S s_total-* | *– lead l*<br>*– spin degree of freedom = 1..m*<br>*- spin degree of freedom = 1..m*<br>*- result of knitting; pointer on "kniter"*<br>*- the system* |
| *return* | *COMPLEX[]* | *A matrix of the size m x m, where m = s_total. lead_size(lead)* |

## 7.4 Module system.py

This module defines functions for visualisation of results

## E. Visualisation of results

| | | |
|---|---|---|
| **F:** | visualisation_simple(unsysteme a) | Makes a simple 2D plot of the system's geometry with black circles (see visu2D) and saves it to file image_simple.eps |
| **args* | *unsysteme a* | *a system to plot;* |
| **F:** | visu2D(unsysteme a, string s) | Makes a 2D plot of the system a: its geometry. Saves the image to file <s>.pdf and <s>.eps. (See also |
| | | visualisation_simple) |
| **args* | *unsysteme a*<br>*string s* | *- system to plot;*<br>*- resulting file;* |
| **F:** | visu3D(unsysteme a, string s) | Makes a 3D plot of the system a: its geometry. Saves the image to file <s>.pdf and <s>.eps. (See also visu2D) |
| **args* | *unsysteme a*<br>*string s* | *- system to plot;*<br>*- resulting file;* |
| **F:** | saveplot2D(unsysteme system, string s_file, <function that returns float>myFun) | Saves the system with associated values at each system point calculated using the function myFun. Resultinf file – <s_file>.data |
| **args* | *unsysteme system-*<br>*string s_file*<br>*<function that returns float> myFun* | *-a pointer to the system to plot,*<br>*-the resulting file name*<br>*-the target function* |
| **F:** | colorplot2D(unsysteme system, string s_file, <function that returns float>myFun) | Plots the system and fills the system area with colors. The color at each system point is calculated based on the value of the function myFun, calculated at this point. |
| **args* | *unsysteme system-*<br>*string s_file*<br>*<function that returns float> myFun* | *-a pointer to the system to plot,*<br>*-the resulting file name*<br>*-the target function that calculates the color values to fill the plot area* |

# 8 System architecture

The KNIT system is based on a C++ library for calculating Green functions in non-interacting tight-biding Models (KNIT - Kernel). It is based on a generalization of the standard Recursive Green Function Algorithm. The library provides, for a very large class of systems, convenient tools to:

1. build tight biding models for multi-terminal systems with complicated geometries and arbitrary internal degrees of freedom (like spin or electron/hole or s,p,d orbitals...)
2. calculate the transport properties of the system (conductance, shot noise, transmission eigenvalues)
3. calculate (out of equilibrium) local quantities inside the system such as local currents or local density of states.

The library is written in C++ and can be used in C++ programs. However, a much more convenient way of using it is through its PYTHON wrapping. PYTHON (a simple and very powerful programming language) allows for a very flexible use of developed libraries for physical calculations and offers a set of intuitive and simple functions for working with graphics.

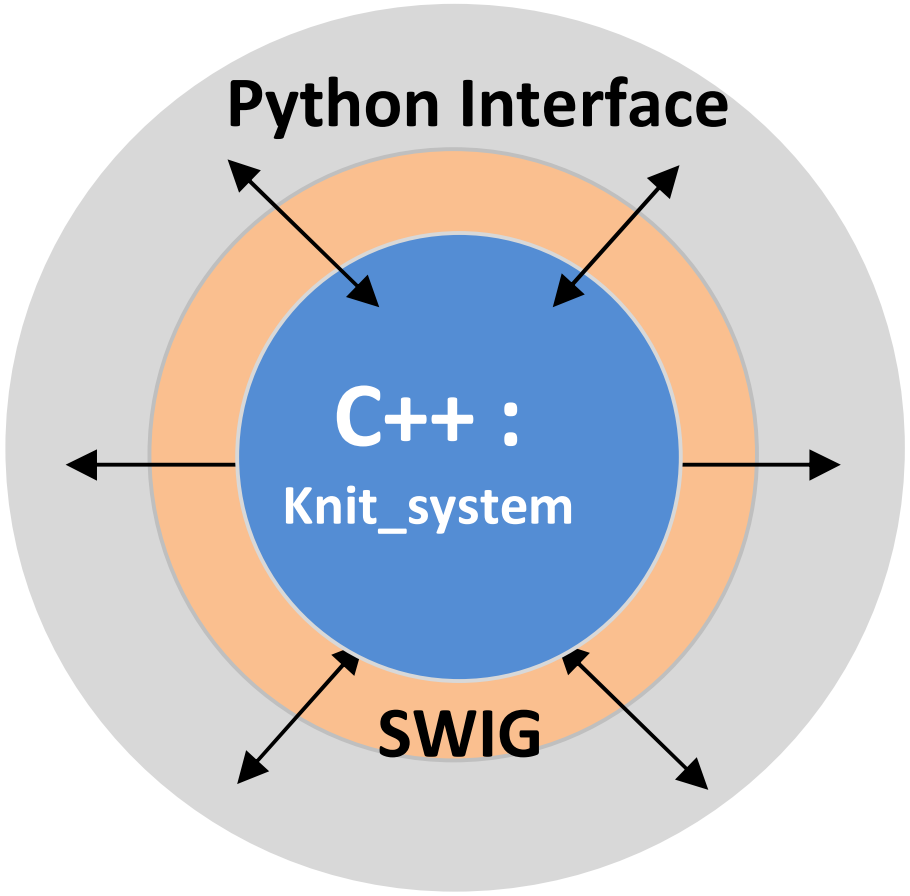

**FIG.9:**

User's commands written in Python specify calls for both native Python functions and C++ functions of the KNIT core. The latter is done by using SWIG. SWIG is a powerful tool for bridging the gap between Python and C++ languages. Compiling the C++ modules with SWIG, Python interfaces for all C++ functions are automatically generated, allowing a modeler to use these functions seamlessly within Python specifications.

6.1 Class diagram

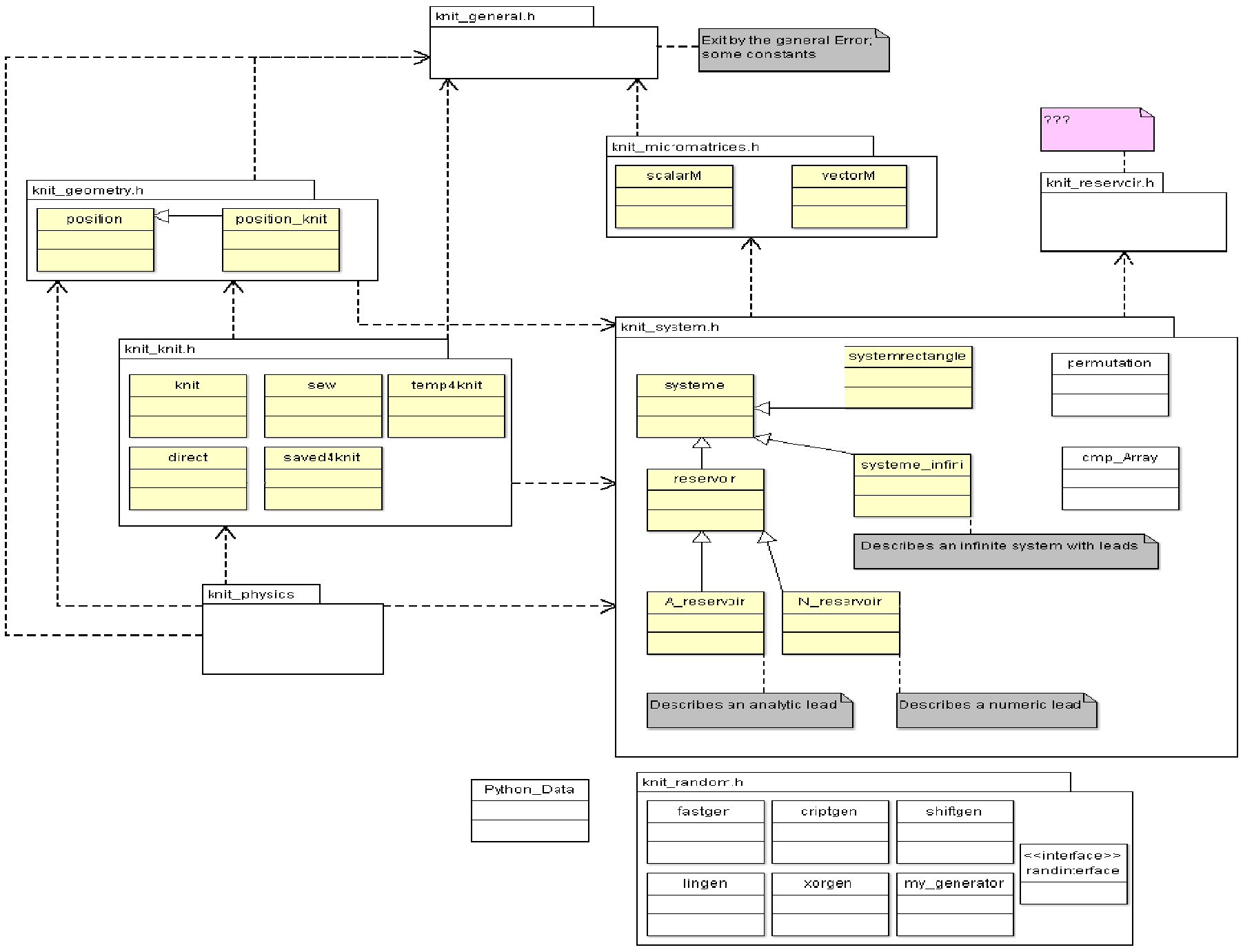